\documentclass[12pt]{article}
\input epsf.tex
\usepackage{amssymb}

\newcommand{\bmat}{\left(\begin{array}}
\newcommand{\emat}{\end{array}\right)}

\def\diagram#1{{\normallineskip=8pt\normalbaselineskip=0pt \matrix{#1}}}

\def\harr#1#2{\smash{\mathop{\hbox to .3in{\rightarrowfill}}
 \limits^{\scriptstyle#1}_{\scriptstyle#2}}}

\def\yzero{\smash{\hbox{$y\kern-4pt\raise1pt\hbox{${}^\circ$}$}}}
\def\ov{\overline}
\def\s2{\frac{1}{\sqrt2}}

\def\beq{\begin{equation}}
\def\eeq{\end{equation}}
\def\beqa{\begin{eqnarray}}
\def\eeqa{\end{eqnarray}}

\def\Dsl{\,\raise.15ex\hbox{/}\mkern-13.5mu D} 

\def\IR{\mathbb{R}}
\def\IC\mathbb{C}
\def\IZ{\mathbb{Z}}
\def\ItZ{\widetilde{\mathbb{Z}}}
\def\IZdos{{\bf Z}_2}

\def\IT{{\bf T}}
\def\IS{\mathbb{S}}
\def\S{{\bf S}}

\def\IRP{\mathbb{R}{\rm P}}

\def\IRPthree{\mathbb{R}{\rm P}^3}
\def\IRPtwo{\mathbb{R}{\rm P}^2}

\def\Dseven{${\widehat{{\rm D}7}}\:$}

\def\Deight{${\widehat{{\rm D}8}}\:$}
\def\Dzero{${\widehat{{\rm D}0}}\:$}
\def\Dthree{${\widehat{{\rm D}3}}\:$}
\def\Dfour{${\widehat{{\rm D}4}}\:$}
\def\Dfive{${\widehat{{\rm D}5}}\:$}
\def\Done{${\widehat{{\rm D}1}}\:$}
\def\Dmone{${\widehat{{\rm D}(-1)}}\:$}
\def\Dtwo{${\widehat{{\rm D}2}}\:$}

\def\Dbseven{${\ov{{\rm D}7}}$}

\def\Dbnine{${\ov{{\rm D}9}}$}
\def\Dbthree{${\ov{{\rm D}3}}$}

\def\Dbeight{${\ov{{\rm D}8}}$}

\def\RROp{$\widetilde{Op}^{\pm}$}

\def\RROfive{$\widetilde{O5}^{\pm}$}

\def\RPOp{$\widetilde{Op}^{+}\:$}

\def\RPOfive{$\widetilde{O5}^{+}\:$}

\def\RPOthree{$\widetilde{O3}^{+}\:$}
\def\RPOtwo{$\widetilde{O2}^{+}\:$}
\def\RPOone{$\widetilde{O1}^{+}\:$}
\def\RPOzero{$\widetilde{O0}^{+}\:$}

\def\RNOp{$\widetilde{Op}^{-}\:$}

\def\RNOseven{$\widetilde{O7}^{-}\:$}
\def\RNOsix{$\widetilde{O6}^{-}\:$}
\def\RNOfive{$\widetilde{O5}^{-}\:$}

\def\RNOtwo{$\widetilde{O2}^{-}\:$}
\def\RNOone{$\widetilde{O1}^{-}\:$}
\def\RNOzero{$\widetilde{O0}^{-}\:$}

\def\RRnsOp{$\widetilde{Op}$}

\topmargin
-1.5cm
\textwidth
15.5cm
\textheight
24cm
\oddsidemargin
0.7cm
\evensidemargin
1.2cm

\begin{document}

\makeatletter
\@addtoreset{equation}{section}
\makeatother
\renewcommand{\theequation}{\thesection.\arabic{equation}}
\pagestyle{empty}
\rightline{CINVESTAV-FIS/02-027}

\rightline{\tt hep-th/0206183}
\vspace{.5cm}
\begin{center}
\Large{\bf Branes and Fluxes in Orientifolds and K-theory}\\

\large
Hugo Garc\'{\i}a-Compe\'an\footnote{ICTP Associate Member, The Abdus Salam International Center
for Theoretical Physics, Trieste, Italy. E-mail address: {\tt compean@fis.cinvestav.mx}} and Oscar
Loaiza-Brito\footnote{Current e-mail address: {\tt loaiza@th.physik.uni-bonn.de}} \\[2mm]
{\em Departamento de F\'{\i}sica}\\
{\em Centro de Investigaci\'on y de Estudios Avanzados del IPN}\\
{\em Apdo. Postal 14-740, 07000, M\'exico D.F., M\'exico}\\ [4mm]
\vspace*{1cm}
\small{\bf Abstract} \\[7mm]
\end{center}

\begin{center} 
\begin{minipage}[h]{14.0cm} 
{RR fields in string backgrounds including orientifold planes and branes on top of them are classified
by K-theory. Following the idea introduced in hep-th/0103183, we also classify such fluxes by
cohomology.  Both of them are compared through the Atiyah-Hirzebruch Spectral Sequence. Some new
correlations between branes on orientifold planes $Op^\pm$ and obstructions to the existence of some
branes are found. Finally, we find a topological condition that avoid the presence of global gauge
anomalies in lower dimensional systems.} 
\end{minipage} 
\end{center}

\vspace{6cm}
  

\leftline{May 2004}

\newpage
\setcounter{page}{1}
\pagestyle{plain}
\renewcommand{\thefootnote}{\arabic{footnote}}
\setcounter{footnote}{0}
\section{Introduction}
String theory backgrounds including orientifold planes have been studied in detail (see
\cite{openstrings,orientifolds,wittenbaryons,hananyo,bgs,sugimoto,orientifolds2, braneboxes, troost}).
Many features of supersymmetric and non-supersymmetric gauge theories have been understood since the
introduction of orientifolds as perturbative and nonperturbative string backgrounds. One of them is the
existence of many types of orientifolds, which arise when discrete fluxes are turned on. The existence
of these fluxes associated to NS-NS and R-R sectors of the theory, change the charge and tension of
orientifold planes. For instance, some of them carry half-integer values of RR charge violating the
Dirac quantization condition. In other cases, planes with the same dimensionality have a relative
charge differing by one half from each other, in D-brane units of charge.

There are at least two different (but related) ways to turn on such discrete fluxes. One of them uses
the fact that branes can end on branes giving a ``brane realization of discrete torsion" (for more
details see section two). The second one is the classification of orientifolds provided by cohomology.
In fact, cohomology groups of the transversal space to orientifolds, classify RR fluxes in the bulk
space. It turns out that some of these fluxes are actually discrete torsion, which in turn, describe
the existence of a new type of orientifold plane. However, cohomology, in general {\it do not} provides
a satisfactory explanation of why some of these orientifolds have a fractional relative charge and
moreover, why some of them (actually those with a spatial dimension less than 5) have indeed a
fractional RR charge.

The problem of the relative charge among some orientifold planes, is successfully resolved by K-theory
(although the problem of the fractional charge for a single orientifold plane is still open). K-theory
has been proved to be a very fruitful mathematical tool to classify D-branes in string theory (see
\cite{wittenk,k-theory,moorek}). Originally, K-theory was used to classify RR charges in different
backgrounds \cite{horova,compean,realk,gukov,branesk}. Recently it was proved that K-theory also
classifies RR fields \cite{moore-witten}, i.e., the fields related to the D-branes at points far away
from orientifold planes (for a formal treatment, see \cite{freed}). Using this result in Ref.
\cite{bgs}, it was possible to classify RR fields in the presence of orientifold backgrounds as well
(with no extra D-branes). Since some of these fluxes turn out to be discrete torsion in the presence of
orientifold planes, such a classification is also an orientifold classification, this time provided by
K-theory.  In fact, comparing the cohomology and K-theory results, it was possible to explain the
relative RR charge among some orientifolds and moreover, new features were found, such as the absence
of certain orientifolds as well as the equivalence between other ones (i.e., some orientifold planes
seemed to be different in the former cohomology classification but they turn out to be the same object
in the K-theory perspective).

In this paper, we are interested in classifying RR fields by K-theory in string theory backgrounds
including orientifold planes and $d$-branes\footnote{Throughout this paper, we are using the following
notation: $d$-branes stands for a $d$-dimensional brane on top of an $Op$-plane for which we do not
know neither its charge nor its nature (i.e., if it is a Dirichlet brane or another type of brane).}.
In particular, we consider the case of $d$-branes on top of the orientifold planes, i.e., only those
oriented parallel to the orientifolds. The orientifold planes and some of these branes can be regarded
as the T-dual versions of the D-branes in Type I and Type $USp(32)$ \cite{usp32} string theories (which
have an $O9^-$ and an $O9^+$-planes respectively) when T-duality is taken over their longitudinal
coordinates (if the number of compact coordinates is higher than the dimensionality of the D-brane, its
T-dual version will be a D-brane transversal to the orientifold plane; such branes are not considered
in the present paper). Also we find new correlations between RR fields in the presence of the branes on
top of orientifolds. In order to do this, we require the knowledge of a cohomology classification. This
give us an alternative method to classify orientifolds by cohomology, when the dimension of the brane
is equal to that of the orientifold. The method consist in wrapping D$(d+n)$-branes on $n$-cycles of
homology to get D$d$-branes. There are certain restrictions in which branes can be wrapped as well as
which cycles are considered, but once we fixed the homology cycles, we are able to compute the
corresponding homology group which classify them. By Poincar\'e duality we get the required cohomology
group. It is important to point out that our results are in agreement with the above mentioned
cohomology classification of orientifolds.

By comparing both results (K-theory and cohomology) we obtain some new correlations among RR fields and
branes. Such a comparison is made by using the Atiyah-Hirzebruch Spectral Sequence (AHSS). Among other
important results, we find that K-theory fixes the topological conditions to cancel global anomalies
arising in probe branes within the same backgrounds we are considering in this paper.

The paper is organized as follows: in section 2 we briefly survey some important aspects of
orientifolds. In section 3 we review how to calculate RR charges of branes on top of an $Op^\pm-$plane.  
Here we describe the T-dual version in which we restrict our study throughout this paper. Also we
discuss on discrete charge cancellation on compact spaces that are reflected on global gauge anomaly
cancellation on suitable probe branes. The \Dthree-brane in Type $USp(32)$ string theory is discussed.

In section 4 we begin by reviewing the classification of RR fields through K-theory and describing how
the AHSS works by relating cohomology to K-theory.  After that we give the relevant K-theory group
which classifies RR fields in orientifolds and $d$-branes backgrounds.

In section 5, we show how to obtain possible $d$-branes on top of orientifold $p$-planes by wrapping
D$(d+n)$-branes on non-trivial and compact $n$-cycles on the projective spaces $\IRP^{8-p}$.

In section 6 we apply the AHSS to relate the results given by K-theory classification of RR fields, and
those given by cohomology.  We interpret the results in the spirit of Ref. \cite{bgs}. This is done for
all $d<p$ and for $0<p\leq 6$. Finally we give our conclusions in section 7. Also in the appendix A we
give detail aspects of transforming fluxes into branes. Some important remarks about T-duality on such
branes are considered in appendix B.

\section{Overview on Orientifolds}

In this section we review some important aspects about orientifold planes (see for instance
\cite{orientifolds,wittenbaryons,hananyo,bgs,sugimoto,braneboxes}).  Our aim is not to provide an
extensive review of orientifolds but to briefly recall some of their relevant properties.

An orientifold plane in Type II superstring theory is defined as the plane conformed by
the loci of fixed points under the action of a discrete symmetry ${\cal I}_{9-p}$, which
reverses the transverse $(9-p)$ coordinates, and that of $\Omega$ which reverses the 
string
worldsheet orientation.  Hence an orientifold $Op$ is given by the plane 
\beqa 
\IR^{p+1}
\times \left( \IR^{9-p}\biggm/ \Omega \cdot {\cal I}_{9-p}\cdot J \right), 
\eeqa 
with $J$ given by (see \cite{sorientifolds}) \beqa J= \left\{ 1 \hfill\hbox{ $p$ = 0,1 mod 4}
\atop (-1)^{F_L} \quad \hbox{$p$ = 2,3 mod 4} \right.\;. \eeqa

There are at least two different types of orientifold planes, denoted as $Op^\pm$, where $\pm$ stands
for the sign of the RR charge they carry on. Actually, they carry a RR charge equal to $\pm 2^{p-5}$ in
D-brane charge units (notice that for $p<5$, the orientifold plane has a fractional charge). These two
different types of orientifold planes can be regarded as arising (via T-duality) from the
nine-dimensional orientifold planes $O9^\pm$, which in turn, establish the existence of the
ten-dimensional string theories, known as type I ($O9^-$) and type $USp(32)$ ($O9^+$) theories. In the
former one, 16 D9-branes (and their images) are needed in order to cancel the $-16$ charge (in D-brane
charge units) due to the orientifold $O9^-$. The physical states (coming from the quantized open
string) are those which survive the action of the projection operator $\widehat{P}_{\Omega} =
\frac{1}{2} (1 + \widehat{\Omega} )$, i.e., the surviving states have an eigenvalue equal to one under
the $\widehat{\Omega}$ action, which acts on the Chan-Paton factors as
\beqa 
\widehat{\Omega} \lambda
|\Psi \rangle = \gamma^{-1}_\Omega\lambda^T\gamma_\Omega |\Omega\Psi \rangle, 
\eeqa 
with $\lambda \in SO(n)$ and $\gamma_\Omega$ being the generating element of a
representation of the $\IZdos$ acting on the Chan-Paton labels, satisfying
$\gamma_\Omega=-\gamma^T_\Omega$. On the other hand, for states satisfying $\gamma_\Omega =\gamma^T_\Omega$
($\widehat{\Omega}^2=1$) the gauge group is $USp(n)$ and the RR charge of the 
orientifold
plane is positive (+16). By taking this orientifold plane, it is possible to construct
the so called $USp(32)$ string theory\footnote{In the usual context, the symplectic
group appears by imposing the conditions $\widehat{\Omega}^2 = 1$, physical 
states with 
eigenvalue $\Omega=1$ and $\gamma_{\Omega}=-\gamma^T_{\Omega}$. However, alternatively  
we can impose
$\widehat{\Omega}^2=1$, physical states with eigenvalue $\Omega=-1$ and
$\gamma_{\Omega}=\gamma^T_{\Omega}$, obtaining also the  symplectic gauge group. 
Tadpole cancellation condition fixes the range of the gauge group to be $32$. This give rise to the 
$USp(32)$
string theory with one $O9^+$-plane. Since the $U(1)$ gauge boson, present in the 
spectrum of
a single D-brane and the NS $B$-field,  are both odd under the orientifold projection, the
difference between Type I and $USp(32)$ string theories lies in the fact that in the former 
one the fields are projected out while in the latter one they are not. The two orientifolds 
$O9^\pm$ differ from
each other by the presence of a non-trivial NS-NS two-form.}, which was proposed in
\cite{usp32}. The above can be summarized as follows:
\begin{eqnarray*} 
\begin{array}{lcr} 
\hbox{Type I} &:& 32\; \rm{D}9 + O9^-
+\hbox{IIB}\\ \hbox{Type } $USp(32)$ &:& 32 \;\hbox{\Dbnine} + O9^+ +\hbox{IIB}.
\end{array} 
\end{eqnarray*} 
By taking T-duality on $(9-p)$ spatial coordinates on the orientifolds $O9^\pm$ (i.e. in Type I and
$USp(32)$ theories), we get the $Op^\pm$ orientifolds (actually $2^{(9-p)}$ of them) and also T-dual
versions of D-branes in the above two ten-dimensional string theories. We focus our attention in
D$d$-branes on top of $Op^\pm$ planes (i.e., D-branes with only longitudinal coordinates with respect to
the orientifold plane). They come from D$(d+9-p)$-branes in the ten-dimensional theories which are
wrapped on the compact $(9-p)$-coordinates. Also, as we know from K-theory, the D-branes present in both
ten-dimensional string theories are D9, D5, D1 (BPS states) and \Deight and \Dseven (non-BPS states) for
Type I string theory, while for Type $USp(32)$ the difference lies on the non-BPS spectrum of branes,
which in this case is given by \Dfour and \Dthree branes.

On the other hand, an orientifold classification can also be provided by a non-perturbative analysis.
This classification is given by cohomology and by turning on discrete fluxes. Before of reviewing this
classification let us start by the analysis of discrete NS fluxes. The transverse space to $Op$ (actually
the projective space $\IRP^{8-p}$) contains a set of non-trivial homology cycles where D-branes or
NS5-branes can be wrapped on. Actually, this picture is ``the brane realization of discrete torsion'',
where an $Op^+$-plane can be constructed by the intersection of an $Op^-$ and a NS5-brane. Hence, it is
important to study the action of an $Op^-$-plane on the $B$-field (for which the NS5-brane is the
magnetic source). It turns out that $B$ is odd under the orientifold projection, which means that $H=dB$
is classified by a torsion cohomology group\footnote{Roughly speaking, a torsion cohomology group,
classifies sections of the bundle $\Omega^3 \otimes {\cal E}$ where ${\cal E}$ is the non-oriented 
line
bundle over $\IRP^{8-p}$, and $\Omega^3$ is the group of three-forms.}. 
Hence, $[H_{NS}=dB_{NS}] \in H^3(\IR{\rm P}^{8-p}, \ItZ)=\IZdos$, with $\ItZ$
being the twisted sheaf-bundle of integers \cite{hananyo}. The trivial class of the two-torsion discrete
group stands for the presence of an $Op^-$ while the non-trivial one is related to the $Op^+$. This can
be understood as follows.

The $B$-field has a non-trivial holonomy given by
\beqa
b\;=\oint_{\IRPtwo}\frac{B}{2\pi}=\frac{1}{2}
\eeqa
with $\IRPtwo \subset \IRP^{8-p}$ surrounding the $Op$-plane. This holonomy
contributes by a factor 
\beqa
g= e^{i\int_{\IRPtwo}B}=e^{i\pi}=-1
\eeqa
to the M\"obius strip amplitude ${\cal M}_2$. Suppose we
start
with an
$Op^--$plane, hence
\beqa
{\cal M}_2 \sim Tr\; \frac{1}{4} \Omega g (1+ (-1)^F)e^{-Ht}.
\eeqa
Then instead of having states invariant under $\frac{1}{2} (1+\widehat{\Omega})$ they are invariant
under $\frac{1}{2} (1- \widehat{\Omega})$.  This means that we have a positive $Op^+-$plane
(According to the footnote in the page 4). So, the presence of a discrete torsion $B$-field produces
the interchange: $Op^- \longleftrightarrow Op^+$.

Let us turn our attention to the discrete RR fluxes. 
Orientifold planes establish an action on the RR $p'$-forms in spacetime. It is
important to know how this action affects the fields. The $B_{NS}$-field
in Type I or $USp(32)$ theory, changes its sign under the action of $O9^\pm$ and it
remains valid for $Op^{\pm}$ with other values of $p$ \footnote{Throughout this paper $p$ stands for
the dimensionality of the 
orientifold plane.}. However for RR
$p'-$forms, the action depends on the dimension of the orientifold, i.e.,
\beqa
\hbox{untwisted}: \: C_{p'} \rightarrow C_{p'} \:\:  \ \ p' = p+1 \; mod\,4\nonumber\\
\hbox{twisted}: \: C_{p'} \rightarrow -C_{p'} \:\: \ \ p' = p+3 \; mod\,4\;.
\label{un-twisted}
\eeqa
There are other kind of orientifolds \cite{wittenbaryons,hananyo,bgs,sugimoto}
given by cohomology torsion variants. Forms of an appropriate rank are
topologically classified by torsion cohomologies, i.e. $[G_{p'+1}] \in
H^{p'+1}(\IR{\rm P}^{8-p}, \IZ)$, where $G_{p'+1} = dC_{p'}$ (field strength
$(p'+1)$-form). Twisted forms given by (2.7) are classified by twisted
cohomologies: $H^{p'+1}(\IR{\rm P}^{8-p} , \ItZ )$. 
For $p\le 6$ there are torsion RR fields which are given by
\beqa
[G_{6-p}] \in H^{6-p}(\IR{\rm P}^{8-p},\IZ (\hbox{or } \ItZ)) = \IZdos.
\label{g}
\eeqa
These are background RR discrete fields and they change some properties of
orientifold planes. 

The main point to focus here is that also (at the cohomological level) RR fields
have torsion, as shown in Eq. (\ref{g}).
Again there is a non-trivial holonomy factor (for $p \le 5$ and besides the trivial
one) given by
\beqa
c\;=\oint_{\IRP^{5-p}}\frac{C_{5-p}}{2\pi}=\frac{1}{2},
\eeqa
that give rise to other kind of orientifold plane denoted by \RRnsOp . So we have
four types of $Op-$planes, according to the holonomies $(b,c)$. The
$(0,0)$-holonomy represents an $Op^-$-plane. $(0,1)$ holonomy is an \RNOp-plane,
$(1,0)$ is an $Op^+$-plane and finally a $(1,1)$ is an \RPOp-plane.

Gauge groups are $USp(2n)$ for the $Op^+$- and \RPOp-planes, although they differ by their dyon
spectrum \cite{hananyo}. For $Op^-$-plane the gauge group is $SO(2n)$ and for \RNOp-plane is
$SO(2n+1)$. By gauge theories and dualities (like the Olive-Montonen duality
\cite{wittenbaryons,hananyo,bgs}) it is known that an \RNOp-plane can be thought as the
configuration $Op^- + \frac{1}{2}Dp$, where $\frac{1}{2}Dp$ is a fractional (stuck) D$p$-brane.

There are extra variants orientifolds $\widehat{Op}$ given by fluxes characterized by cohomology
groups $H^{2-p}$ and they are valid only for $p<2$.

However there are more restrictions. For example, it was shown in Ref. \cite{sugimoto} that
\RRnsOp-planes do not exist for $p\geq 6$, with the exception of the \RNOsix-plane, which can be
realized as an $O6^-$ immersed in a non-zero background cosmological constant (massive Type IIA
supergravity; see section 6.2).  Also, we have learned from \cite{bgs} (see section 4 and 5 for
details) that $Op^+$ and \RPOp, for $p\le 3$, are equivalent in K-theory, and moreover,
$\widehat{Op}^+$ and $\widetilde{\widehat{Op}}^+,$ do not exist for $p < 2$.

As it was said, turning on discrete fluxes, they can be studied as brane realizations. The $b$ holonomy
factor is obtained by intersecting $NS$5-branes and $Op$-planes, while the $c$ holonomy factor is
obtained by intersecting D$(p+2)$-branes and $Op$-planes. Readers interested in the details of these
issues are invited to consult Ref. \cite{braneboxes} (see also \cite{hananyo,bgs}).


\section{D$d$-branes on Top of $Op^\pm-$planes}

D-branes in type I theory are classified by {\it real} K-theory\footnote{The action of the worldsheet
parity $\Omega$ induces an antilinear involution on the gauge bundles $E$ over $X$ that commutes with
$\tau$, where $\tau$ is the involution $\tau: X \rightarrow X$, $\tau^2 =id$, related to the inversion
of transverse coordinates to $Op$. The Grothendieck group of the isomorphism classes of these bundles
is called the {\it real} K-theory, and it is denoted by $KR(X)$.} \cite{gukov, realk, wittenk,
branesk}, while those in type $USp(32)$, are classified by quaternionic K-theory (see below). If we
apply T-duality on $(9-p)$ compact directions in the above theories, we get a spectrum of D-branes,
which are parallel or transversal, to $Op$-planes. It turns out that real and quaternionic K-theory
still classifies such D-branes in the presence of lower dimensional orientifolds. In this section we
briefly review how to classify D$d$-branes on top of $Op$-planes, specifically, the ones we are
interested in this paper, which are parallelly oriented to the orientifold plane (which in turn means
that $d\leq p$). In the second part of this section, we study global gauge anomalies, arising on
suitable probe branes on compact spaces, due to the presence of both kinds of orientifold planes,
$Op^\pm$. Our interest in these anomalies, lies in the fact that we will be able to predict the
suitable conditions to cancel them, by using K-theory (see section 5).

\subsection{K-theory classification of D$d$-branes on orientifolds $Op^\pm$}
Before of describing how real K-theory\footnote{For mathematical
properties of real K-theory see \cite{realk} and for physical applications
see Refs. \cite{branesk,bergmant,bergmanline, wittenk}.} is used to compute RR charges in string
theory (with no $B$-field in the background) it is useful to give the main
properties of these K-theory groups. Consider the following definition $\IR^{p,q} := (\IR^p
/ \IZdos) \times \IR^q$ where $\IZdos$ inverts $p$ coordinates.
$\IS^{p,q}$ is defined as the unitary sphere in $\IR^{p,q}$ with dimension
$p+q-1$. Then, $\IS^{p,0} \cong \IRP^{p-1}$ and $\IS^{0,q} \cong \S^{q-1}$.

Real K-theory groups satisfy the following properties:
\beqa
\begin{array}{ccl}
KR^{-n}(X)&=& KR^{0,n}(X),\nonumber\\
KR^{p,q}(X)&=& KR(X \times \IR^{p,q}),\nonumber\\
KR^{p,q}(X)&=& KR^{p+1, q+1}(X) =KR^{p-q}(X),\\
KR^{-m}(X)&=&KR^{-m-8}(X).\nonumber
\end{array}
\eeqa
The same relations are valid for the quaternionic case with $KR^{-n}(X)\cong KH^{-n+4}(X)$. In
order
to give a complete classification of RR charges on orientifold backgrounds by
K-theory, let us describe briefly some results given in
Refs. \cite{bgs,wittenk,gukov,bergmant}.

It was proposed in \cite{wittenk} that real K-theory classifies RR charges of
D$d$-branes on top of $Op^-$-planes, while quaternionic K-theory make the same for
branes on $Op^+$-planes. In \cite{gukov} this computation was done explicitly.

Also, it was shown in \cite{bergmant} that the K-theory group classifying RR
charges in  type I T-dual models\footnote{Actually T-duality acts on derived categories, or roughly speaking, on 
K-theory. See Ref. \cite{indext}.} is the relative group
\beqa
KR^{p-9}(\S^{p-d}\times\IT^{9-p},\IT^{9-p})\cong \bigoplus^{9-p}_{s=0} {9-p \choose
s}KO^{-s}(\S^{p-d}),
\label{bergmant}
\eeqa
with $(9-p)$ being the number of coordinates under which  T-duality acts.
A similar formula holds for T-dual models of $USp(32)$ string theory, with quaternionic
groups.

The groups on the right hand side of (\ref{bergmant}) with $s \neq 9-p$ classify charges for
wrapped D$(d+9-p)$-branes in the ten dimensional theory (with $s$ being the number of
wrapped coordinates) and those with $s=9-p$ classify unwrapped D$d$-branes in ten
dimensions on an $Op-$plane.

However, we are interested just in D$d$-branes obtained by wrapping 
D$(d+9-p)$-branes (either in Type I or $USp(32)$  string theory) on coordinates
$d+1,\cdots , 9-p$ (notice that we are not interested in branes with transversal coordinates to $Op$). 
The real K-theory group classifying these kind of fluxes is given by,
\beqa
KR^{p-9}(\IR^{9-p,p-d})\cong KO(\S^{p-d}),
\label{cargas1}
\eeqa
and it is valid for all $p$.
Hence, this is the relevant group that classify wrapped D$(d+9-p)$-branes, in the ten
dimensional theory, on $\IT^{9-p}$ and on top of an $Op$-plane. From now on, we will
refer to these
branes as the T-dual version of branes on Type I (or $USp(32)$ ) theories until we
require to be more specific.
For $Op^+$-planes, T-dual version of branes in $USp(32)$ string theory,
are given by the quaternionic K-theory group,
\beqa
KH^{p-9}(\IR^{9-p,p-d})\cong KSp(\S^{p-d})\cong KO(\S^{p-d+4}).
\label{cargas2}
\eeqa
In the next section  we will compute RR fields associated to these kind of branes.

\subsection{Discrete charges and global gauge anomalies}

Anomalies in probe branes on compact spaces are related to non-zero RR discrete charges
and with the presence of \RROp-planes for $p>6$
\cite{urangak,sugimoto}.

We are interested on global gauge anomalies \cite{su2} arising in intersecting probe D-branes
with discrete charge on compact spaces. In \cite{urangak} it
was shown that by using D5 probe branes wrapped on \IT$^2$ in Type I theory, that 
\Dseven -branes should exist in an even number in order that
global gauge anomalies be canceled, i.e., discrete K-theory charge should be canceled.

The idea is to consider $2n$ coincident D5-branes wrapped on \IT$^2$ and one \Dseven sitting at a point
in \IT$^2$. The four dimensional non-compact space that intersects the \Dseven -brane contains fields
arising from strings attached to both branes. Since \Dseven = D7 +\Dbseven /$\Omega$ from IIB theory,
it is enough to compute the sector 75 and $\bar{7}5$. The result is the existence of a four-dimensional
Weyl fermion in the fundamental representation $2n$ of $USp(2n)$. This gives rise to a $SU(2)$ global
gauge anomaly \cite{su2}. The argument can be extended to orientifolds {\bf T}$^4/\IZdos$ in the IIB
theory.

Now we want to show that also \Dfour - and \Dthree -branes in $USp(32)$ string theory give rise to
global gauge anomalies on suitable probe D-branes. We consider a compactification of $USp(32)$ theory
on \IT$^6$ with a single \Dthree -brane extending along the four non-compact dimensions, and placing it
at a point in \IT$^6$. These systems contain tachyonic modes arising from the $\widehat{3}\bar{9}$ and
$\bar{9}\widehat{3}$ open string sectors. The \Dthree -brane carries a $\IZdos$-charge measured by
K-theory. In this case, the suitable probe branes are the $\bar{9}$-branes themselves (remember that
there are \Dbnine -branes because tadpole cancellation in $USp(32)$ string theory). \Dthree -brane is
constructed in string theory as a Type IIB D3-\Dbthree -pair exchanged by $\Omega$. Let us compute the
nonsupersymmetric spectrum arising from $\widehat{3}\bar{9}$ and $\bar{9}\widehat{3}$ sectors. Sectors
$\bar{9}3$ and $\bar{9}\bar{3}$ are mapped into $\bar{3}\bar{9}$ and $3\bar{9}$. In the fermionic
content there is a Weyl fermion in the fundamental representation $2n$ of $USp(2n)$. This is
inconsistent at the quantum level. Thus, the \Dthree -branes should appear in pairs on compact spaces.

We conclude that for $\widehat{D}(p-6)$-branes on top of an $Op^+$-plane (with $p=5,6$) also must be in
pairs. The same result is valid for the \Dfour$\;$ in $USp(32)$ string theory and for
$\widehat{D}(p-5)$ -branes (for $p=4,5,6$) on T-dual versions of $USp(32)$ theory with $Op^+$-planes.


\section{RR Fields, Orientifolds and K-theory}

The aim of this section is to classify RR fields in the presence of orientifold planes and branes on
top of them.  The procedure is as follows: firstly, we give a briefly review of K-theory classification
of RR fields in type II string theories; this survey is based in \cite{moore-witten}. Secondly, we
review the K-theory classification of RR fields with orientifolds, which was studied in \cite{bgs}.
Finally, we take the results given in section 3 and the K-theory classification of RR fields with
orientifolds (given in the present section) in order to obtain the K-theory classification of RR
fields in the presence of orientifold planes and $d$-branes (on top of the $Op$-planes). As it is shown
below, we get a K-theory group which classifies RR fields in such backgrounds.

\subsection{RR fields and K-theory}

It is well know that D-brane charges are classified by K-theory rather than
by cohomology. Recently it was shown that also RR fields are classified
by K-theory  \cite{moore-witten}, even though they are not related to a source. Let us
remind this important fact.

It is possible to show that RR charge is measured by the kernel of the map $i:K({\cal M},{\cal
N};\IZ) \rightarrow K({\cal M};\IZ)$, with $K({\cal M},{\cal N};\IZ)$ being the K-theory group which
classifies classes of bundles on $\cal M$ that are trivial on $\cal N$, where $\cal M$ is the
spacetime manifold and $\cal N$ is its boundary. The important fact is that
\beqa
Ker(i)= K^1({\cal N})/ j(K^1({\cal M})),
\eeqa
with $j$ the restriction to the boundary ${\cal N}=\partial{\cal M},$ 
and where $K^1({\cal N})$ classifies RR fields at infinity and
$K^1({\cal M})$ classifies fields on $\cal M$ that {\it do not have any brane
source} (in Type IIB theory), i.e., the K-theory classification of RR charges is given by the group $K({\cal M})$, while RR fields are
classified by $K^1({\cal M})$.

The result is easily extended to Type IIA and Type I theories. The groups are $K({\cal M})$ and
$KO^{-1}({\cal M})$ respectively.

\subsection{Real K-theory and orientifold classification}

Although we have seen a cohomological classification of orientifold planes, there are some issues
that cohomology is not able to explain. For instance, when discrete fluxes are turned on, the charges
and tensions of orientifold planes are changed, giving rise to different types of planes for the same
dimensionality, as it was seen in section 2. For instance, the charge of $Op^-$ differs with respect
to \RNOp by one half (in D-brane units of charge). This issue is not explained by 
cohomology. However, by using K-theory, Bergman, Gimon and Sugimoto (BGS) \cite{bgs} explained the
relative charge between the above orientifolds, and moreover, they found some new correlations among
other types of orientifold planes. This was done by the K-theory classification of RR fields in the
presence of orientifold planes.

The K-theory groups which classify RR fields in orientifold backgrounds, according to BGS, are given by
\beqa
\begin{array}{cl}
Op^-:& KR^{p-10}(\S^{9-p,0}),\\
Op^+:& KR^{p-6}(\S^{9-p,0})=KH^{p-10}(\S^{9-p,0}).
\end{array}
\label{bgseqs}
\eeqa
They are easily calculated by using the Atiyah isomorphism
\beqa
KR^{-n}(\IS^{p,0}\times X) \, = \, KR^{p-n+1}(X) \oplus KR^{-n}(X),
\eeqa
with $X=\{pt\}$ and by knowing the groups for a point space, which read 
\beqa
KR^{-n}(\{pt\})=\{ \IZ,\IZdos, \IZdos, 0,\IZ,0,0,0\}\; {\rm mod} \; 8.
\eeqa

\noindent
{\bf An example: The $O5$-plane}.
Let us explain some important details of this classification by analyzing one specific example: the
orientifold five-plane. The cohomology classification of this orientifold, as we saw in section 2, is given by the groups
$H^3(\IRP^3, \IZ)=\IZ$ (which give us the integer RR charge of D5-branes on top of it), and
$H^1(\IRP^3;\ItZ)=\IZdos$ (the non-trivial element of $\IZdos$ give us the existence of the
orientifold variant \RROfive). Notice that we are classifying orientifolds according to the
cohomology group of RR fields, that is the reason why we have actually one single group for the two
variants $O5^\pm$, which means that a cohomology of RR forms does not distinguish between $O5^+$ and $O5^-$ planes.
Now, according to the above results, the K-theory classification of orientifold five-planes is given
by the groups $KR^{-5}(\IS^{4,0})=\IZ$ for $O5^-$ and $KH^{-5}(\IS^{4,0})=\IZ\oplus \IZdos$ for
$O5^+$. Notice that these groups are classifying RR fields and that give us a
different result as the RR charges classification (which just give us the value of $\IZ$ for both
cases). So, there are some important questions to address: what does this difference between RR
fields and RR charges K-theory classification mean? and what does the difference between cohomology
and K-theory means? Both of them were correctly answered by BGS. The answer for the first question is
that there are RR fields not related to D-branes but to the presence of orientifold planes, i.e.,
discrete RR fields are turned on by placing orientifold planes in the background. That is the reason
why a K-theory classification of RR fields is in fact, an orientifold classification. The answer for
the second question involves a feature which has been well studied by mathematicians. Indeed, there
is an algorithm which relates cohomology to K-theory and that, also gives some physical consequences
when it is applied to the above case. Hence, before we continue describing the case of the
orientifold five-plane, let us review this algorithm called the Atiyah-Hirzebruch Spectral Sequence
(AHSS).

\subsubsection{The Atiyah-Hirzebruch Spectral Sequence (AHSS)}

The AHSS is an algebraic algorithm that allows to relate K-theory to integral cohomology (see for
instance, \cite{bgs,bergmant,wittend}).

The basic idea of the AHSS is to compute $K(X)$ using a sequence of successive approximations, starting with integral cohomology\footnote{For
an introductory review of the AHSS see \cite{bgs} and references therein. Also see \cite{mms}.}. Basically each step of approximation is given
by the cohomology of a differential operator $d_r$, denoted as
\beqa
E^p_{r+1}=ker\;d_r/Im\;d^{p-r}_r,
\eeqa
where $d^p_r: E^p_r \rightarrow E^{p+r}_r$. In each step, we refine the approximation by removing cohomology classes which are not closed
under the differential $d^p_r$. Closed classes survive the refinement while exact classes are mapped to trivial ones in the next step. In the
complex case (without orientifolds), the first non-trivial higher differential is given by $d_3=Sq^3+H_{NS}$, where $Sq^3$ is the Steenrod
square and $H_{NS}$ is the NS-NS three form. In the case of string theory, the only possible next higher differential is $d_5$.

By the above procedure we get the associated graded complex $GrK(X)$ which is the approximation to
$K(X)$. The graded complex is given by
\beqa
GrK(X)=\oplus_p E^p_r=\oplus_pK_p(X)/K_{p+1}(X)
\eeqa
where $K_n(X)\subset K_{n-1}(X)\subset \cdots\subset K_0(X)=K(X)$. At the first
approximation we have
\beqa
K_p(X)/K_{p+1}(X) = \left\{ H^p(X,\IZ) \quad \hbox{for $p$ even}
\atop                           0 \quad \hbox{for $p$ odd}
\right.
\eeqa
for Type IIA theory, and
\beqa
K_p(X)/K_{p+1}(X) = \left\{ H^p(X,\IZ) \quad \hbox{for $p$ odd}
\atop                           0 \quad \hbox{for $p$ even}
\right.
\eeqa
for Type IIB. Thus, computing $K(X)$ implies that we have to resolve the follow extension
problem,
\beqa
\diagram{
0&\harr{}{}&K_{p+1}(X)&\harr{}{}&K_p(X)&\harr{}{}&K_p(X)/K_{p+1}(X)&\harr{}{}&0\cr
}\;.
\eeqa
If the above sequence is trivial we have that
\beqa
K_p(X)=K_{p+1}(X)\oplus K_p(X)/K_{p+1}(X).
\eeqa
If all extensions are trivial, then $K(X)=GrK(X)$. In our case, we just have to worry
about the mapping $d_3$. If $d_3$ is trivial we finish at the cohomology level, and we must ask about the
exactness of the sequence. When the sequence is not exact, $p$-forms of different degree become
correlated and physically this means that we have correlations between the associated RR fields.

For real K-theory (or in general, for K-theory groups with freely acting involutions) the
approximations are given by {\it twisted} or {\it untwisted} maps (see appendix in \cite{bgs}), i.e., 
the $d_3$
differential operator maps twisted into untwisted classes and vice versa. In this case
$d_3=\widetilde{Sq^3}+H_{NS},$ with $[H_{NS}]\in \IZdos$. It is assumed that $\widetilde{Sq^3}$ is
trivial for both values of $\IZdos$ (i.e., for $Op^+$ and $Op^-$) and $d_5$ is trivial in all cases.
$d_5$ maps (un)twisted into (un)twisted classes.

The first approximation to the graded complex $GrK^{-s}(X)$$=\bigoplus_n E^{p,-(p+s)}_n$,
with
\beqa
E^{p,-(p+s)}_n(X)\,=\,K^{-s}_p(X)/K^{-s}_{p+1}(X),
\eeqa
is given by
\beqa
\begin{array}{ccl}
E^{p,q}_1&=&C^p(X|_{\tau},\IZ)\; \hbox{for $q = 0\;{\rm mod}\;4$}\\
E^{p,q}_1&=&C^p(X|_{\tau},\ItZ)\; \hbox{for $q = 2\;{\rm mod}\;4$}\\
E^{p,q}_1&=&0\;\hbox{for $q$ odd,}
\end{array}
\label{ahss0}
\eeqa
where $\tau$ is the freely acting involution on $X$. Then, the second order of this approximation is 
given by 
the cohomology groups
\beqa
\begin{array}{ccl}
E^{p,q}_2&=&H^p(X|_{\tau},\IZ)\; \hbox{for $q = 0\;{\rm mod}\;4$}\\
E^{p,q}_2&=&H^p(X|_{\tau},\ItZ)\; \hbox{for $q = 2\;{\rm mod}\;4$}\\
E^{p,q}_2&=&0\;\hbox{for $q$ odd}.
\end{array}
\label{ahss}
\eeqa
The same results stand for quaternionic K-theory groups.\\

It is important to point out that triviality of Steenrod square which is also taken in the
untwisted version, actually has a physical interpretation. $Sq^3=0$ implies that
$W_3(Q) =0$, where $W_3$ is the Bockstein homomorphism and the above relation
expresses the fact that a
D-brane can be wrapped on a submanifold $Q$. This means that $Q$ must be a Spin$_c$
manifold \cite{wittenk}. When the NS $H$-field is different from zero, the required topological
condition is
\beqa
[H_{NS}] + W_3(Q)=0.
\eeqa
It is shown in \cite{freed-witten} that this is in fact the condition to cancel
anomalies arising in the worldsheet of strings in the presence of D-branes in
Type II theory. On the other hand, the AHSS described in terms of branes (see
appendix of \cite{wittend}), requires to wrap D-branes on submanifolds $Q$. Thus,
in order to lift trivially cohomology forms to K-theory, we need that $d_3=0$, or
that suitable D-branes wrap on Spin$_c$ manifolds. When this submanifold is not
Spin$_c$ cohomology and K-theory differ from each other.

Now, for the twisted version of $d_3$ it is assumed the same triviality in the
twisted version of the Steenrod square. This means that a topological condition
could be also present for the case of Type I theory and then, there is an
anomaly present in the worldsheet of open strings in the presence of D-branes
and orientifolds of Type II theories. It would be interesting to study what could
be the `twisted' version of a Spin$_c$ manifold.

\noindent
{\bf Example: The $O5$-plane}

Once we have a procedure to compare or lift cohomology to K-theory, and by knowing the K-theory 
groups which classify RR fields related to orientifold
planes, it is possible to get a physical picture which interprets the difference between cohomology and K-theory. Let us come back to our
example of the orientifold five plane. In this case $d_3$ is trivial for both types of $O5$-planes, as well as $d_5$. Hence, the
approximation ends at cohomology. It is possible to show that the extension problem to solve is
\beqa
\diagram{
0&\harr{}{}&\IZ&\harr{\left\{\times 2\hfill\hbox{for $O5^-$}\atop Id\quad\hbox{for $O5^+$}\right\}}{}&\left\{\IZ\hfill\hbox{for $O5^-$}
\atop \IZ\oplus\IZdos\quad\hbox{for $O5^+$}\right\}&\harr{}{}&\IZdos&\harr{}{}&0\cr
 &         &\parallel&&\parallel&&\parallel&&\cr
&&H^3&&\left\{KR^{-5}(\IS^{4,0})\hfill\hbox{for $O5^-$}\atop KH^{-5}(\IS^{4,0})\quad\hbox{for $O5^+$}\right\}&&\widetilde{H^1}&&
}\;.
\eeqa
In the case of the $O5^+$-plane, the sequence is trivial while for the case of $O5^-$ it is not. In the
latter case this means that a half-integer shift is produced in $H^3$ due to the presence of the flux
$G_1\in H^1$. The physical implication is as follows: cohomology gives us a classification of
orientifolds that must be refined by K-theory. The refinement is produced by the half-integer shift in
the flux $G_3$, or in other words, by a half-integer shift in the RR charge of the orientifold $O5^-$.
Afterwards, the K-theory picture, trough the application of the AHSS, explain why the \RNOfive -plane
has precisely, an extra half-integer amount of RR charge than the ordinary $O5^-$-plane. So, an
\RNOfive can be written as $O5^- + \frac{1}{2}\widehat{{\rm D}5}$. The same description holds for all
the lower orientifolds \RNOp. In the case of an $O5^+$ plane there is no an extra shift in the RR
charge of \RPOfive, and then its charge is the same than an $O5^+$-plane. This case is trivial and
cohomology gives an exact description of the K-theory group (the graded complex is equal to the
K-theory group). The anti-D5-brane on top of the $O5^+$-plane corresponds to a stable but
non-supersymmetric system \cite{urangaso}.

Another interesting result involves the $O3$-plane. In such a case, the approximation given by the
AHSS, does not ends at the first step, since $d_3$ is not trivial for $O3^+$ (for $Op^-$, $d_3$ is
always trivial since $H_{NS}=0$ and the twisted version of $Sq^3$ is trivial as well). Hence, the
non-trivial discrete class of $H^3(\IRP^5;\ItZ)$ (which at the cohomology level suggests the presence
of an \RPOthree -plane) is obstructed to be lifted to K-theory (it is not a closed form under $d_3$).
The conclusion is that both orientifolds, $O3^+$ and \RPOthree, are actually the same object.

\subsection{Branes, orientifolds and K-theory}

We have seen that RR fields are classified by K-theory even if they are
source-free. Also that this feature allow us to classify RR-fields in
orientifold backgrounds and  to find some correlations between $Op^\pm-$planes.

Now we are interested in classifying RR fields in the presence of orientifold planes, and $d$-branes,
with $d<p$. We expect to obtain RR fields associated to D$d$-branes (T-dual versions of those
D-branes living on Type I and $USp(32)$ string theories) present on top of an $Op^\pm-$plane, i.e.,
with all their coordinates along the orientifold (they are in the set of RR charges
classified by K-theory). We are also interested in classifying RR fields that are not associated to the
above D-branes, and which in turn be discrete fluxes in the background.

In order to classify these fields, we have to answer first some questions:
\begin{enumerate}
\item
Which is the K-theory group that classifies RR fields in the presence of $d$-branes and
$Op-$planes?
\item
If we want to find charge correlations, as was done in \cite{bgs} for the orientifolds, we
must
use the AHSS. But this requires the knowledge of (related) cohomology groups. Thus,
which are the relevant cohomology groups classifying RR forms with $d$-branes and  $Op$-planes?

\item If there are RR fields without a source in the presence of orientifold and $d$-branes, what is
the role played by $d$-branes associated to such fields?

\end{enumerate}

Let us start by answering the first question. For that, we need to describe how to wrap D8-branes on
spacetime in order to know which K-theory groups are the relevant ones to classify RR fields in the
mentioned conditions.  In Ref. \cite{bgs} a D8-brane was wrapped on a $\S^{8-p}$ sphere on the covering
transverse space $\IR^{9-p}$. After taking the orientifold action, the transverse space is $\IR{\rm
P}^{8-p}$. Hence, actually one is wrapping a D8-brane on $\IS^{9-p,0},$ which is the unitary sphere
on $\IR^{9-p,0}$. By this procedure, BGS get the K-theory groups given in Eq. (\ref{bgseqs}).

Now we want to wrap 8-branes on the transverse space to a $d$-brane on top of an
$Op$-plane, for $d<p$. Then we wrap 8-branes on
\beqa
\S^{8-p}\times \IR^{p-d}
\eeqa
in the covering space. Two comments are in order. First, note that the
space $\IR^{p-d}$ is transverse to the $d$-brane but it is still immersed in the
$Op$-plane. Second, for
$p=d$ we recover BGS results \cite{bgs}. The above product of spaces can be written as
\beqa
\IS^{9-p,0}\times \IR^{0,p-d}.
\eeqa
Now, in order to show that the K-theory groups, which classify RR fields for $Op-$planes,
do not change their order for a fixed value of $p$, in relation to the suitable groups given in \ref{bgseqs},
consider the $O8$-projection
$\IR^1/{\cal I}_1\Omega$ in Type IIA theory. $O8$ maps a D8 wrapping a point on one
side of the orientifold to an image D8 wrapping the other point (looks like a wrapped
\Dbeight ). So the relevant K-theory group is $KR_\pm$ for Type I' (with two $O8^-$) and $KH_\pm$ for
the T-dual version of $USp(32)$ (with two $O8^+$).

Now by wrapping a D8-brane on a $\S^0 \times \IR^1$ (taking for instance $d=7$), the transverse space
to a 7-brane inside an $O8$-plane is actually divided into two parts. Since the fraction of the D8
wrapped on $\IR^1$ is on the orientifold, it is its own self-image since its orientation can be
regarded as the orientation of an anti-brane with reversal orientation. Hence, repeating this
procedure for all $Op$-planes we conclude that K-theory groups must be exactly the same than those 
given by
BGS, but over different suitable spaces. In other words, $\IR^{p-d}$ is fixed under the orientifold
projection.

Hence, K-theory groups that classify RR fields on an orientifold and D$d$-branes
backgrounds are:
\beqa
\begin{array}{cc}
Op^- : \quad & KR^{p-10}(\IS^{9-p,0}\times \IR^{0,p-d}),\\
Op^+ : \quad & KR^{p-6}(\IS^{9-p,0}\times \IR^{0,p-d}).
\end{array}
\eeqa
Using the Atiyah isomorphism \cite{realk}, we get
\beqa
\begin{array}{ccl}
KR^{-n}(\IS^{m,0} \times \IR^{0, l}) &=&\, KR^{-n+m+1}(\IR^{0, l}) \oplus
KR^{-n}(\IR^{0, l})\\
&=&KR^{-n+m+1,l}(\{pt\}) \oplus KR^{-n,l}(\{pt\})\\
&=&KR^{-n+m-l+1}(\{pt\}) \oplus KR^{-(n+l)}(\{pt\})\\
&=&KR^{-(n+l)}(\IS^{m,0}\times \{pt\}).
\end{array}
\eeqa
Replacing the variables by taking
\beqa
\begin{array}{ccc}
-n& \rightarrow &p-10\\
 m& \rightarrow &9-p\\
l&\rightarrow &p-d,
\end{array}
\eeqa
we get our final expression that allow us to calculate RR fluxes on a $d$-dimensional submanifold
within the orientifold $Op^-$, or the RR fluxes related to $d$-branes on top of orientifold
planes. For $Op^+$-planes we have similar results\footnote{The following expressions are just valid
for $p\le 6$. For $p>6$ we have the usual results, i.e., the second term of the right-hand side is not
present since Atiyah isomorphism is not longer valid.}:

\beqa
\begin{array}{cc}
Op^- : & \quad KR^{d-10}(\IS^{9-p,0})\,=\, KR^{d-p}(\{pt\})\oplus KR^{d-10}(\{pt\})\;,\\
Op^+ : & \quad KR^{d-6}(\IS^{9-p,0})\,=\, KR^{d-p+4}(\{pt\})\oplus KR^{d-6}(\{pt\}).
\end{array}
\label{formula}
\eeqa

The results of the computation of these groups for $Op^-$ and $Op^+$ planes are summarized in tables
\ref{Kop-} and \ref{Kop+} respectively. Notice that for the case $d=p$ we recover BGS results
\cite{bgs}. Once we have calculated these groups many interesting issues result from it. In the next
sections we will describe some of them.

\begin{table}
\begin{center}
\caption{RR fluxes for $d$-branes on top of $Op^--$planes.}
\label{Kop-}
\begin{tabular}{||c||c|c|c|c|c|c|c|c|c||} \hline\hline
 $d$&$O8^-$&$O7^-$&$O6^-$&$O5^-$&$O4^-$&$O3^-$&$O2^-$&$O1^-$&$O0^-$\\ \hline\hline
8 &$\IZ$    &        &               & & & & & &  \\ \hline
7 &$\IZdos$ & $\IZ$    &               & & & & & &  \\ \hline
6 &$\IZdos$ & $\IZdos$ & $\IZ \oplus \IZ$ & & & & & &  \\ \hline 
5 & 0 & $\IZdos$ & $\IZdos (\oplus 0)$ & $\IZ (\oplus 0)$ & & & & &  \\ \hline
4 & $\IZ$ & $0$ & $\IZdos (\oplus 0)$&$\IZdos (\oplus 0)$&$\IZ (\oplus 0)$&&&&\\
\hline
3&$0$&$\IZ$&$0$&$\IZdos (\oplus 0)$&$\IZdos (\oplus 0)$&$\IZ (\oplus 0)$&&&\\
\hline
2&$0$&$0$&$\IZ \oplus \IZ$&$(0 \oplus )\IZ$&$\IZdos \oplus \IZ$&$\IZdos \oplus
\IZ$&$\IZ \oplus \IZ$&&\\ \hline
1&0&0&(0 $\oplus )\IZdos$&$\IZ \oplus \IZdos$&(0 $\oplus ) \IZdos$&$\IZdos \oplus
\IZdos$&$\IZdos \oplus \IZdos$&$\IZ \oplus \IZdos$& \\\hline
0&$\IZ$&0&(0 $\oplus )\IZdos$&(0 $\oplus ) \IZdos$&$\IZ \oplus \IZdos$&(0 $\oplus )
\IZdos$&$\IZdos \oplus \IZdos$&$\IZdos \oplus \IZdos$&$\IZ \oplus
\IZdos$\\ \hline
(-1)&$\IZdos$&$\IZ$&0&0&0&$\IZ (\oplus 0)$&0&$\IZdos (\oplus 0)$&$\IZdos (\oplus 0)$\\ \hline\hline
\end{tabular}
\end{center}
\end{table}

\begin{table}
\begin{center}
\caption{RR fluxes for $d$-branes on top of $Op^+-$planes.}
\label{Kop+}
\begin{tabular}{||c||c|c|c|c|c|c|c|c|c||} \hline\hline
$d$ &$O8^+$&$O7^+$&$O6^+$&$O5^+$&$O4^+$&$O3^+$&$O2^+$&$O1^+$&$O0^+$\\ \hline\hline
8&$\IZ$&&&&&&&&\\ \hline
7&0&$\IZ$&&&&&&&\\ \hline
6&0&0&$\IZ \oplus \IZ$&&&&&&\\ \hline
5&0&0&(0 $\oplus )\IZdos$&$\IZ \oplus \IZdos$&&&&&\\ \hline
4&$\IZ$&0&(0 $\oplus )\IZdos$&(0 $\oplus )\IZdos$&$\IZ \oplus \IZdos$&&&&\\ \hline
3&$\IZdos$&$\IZ$&0&0&0&$\IZ$&&&\\ \hline
2&$\IZdos$&$\IZdos$&$\IZ \oplus \IZ$&(0 $\oplus )\IZ$ &(0 $\oplus )\IZ$&(0
$\oplus )\IZ$&$\IZ \oplus \IZ$&&\\ \hline
1&0&$\IZdos$&$\IZdos (\oplus$ 0)&$\IZ (\oplus$ 0)&0&0&0&$\IZ$&\\ \hline
0&$\IZ$&0&$\IZdos (\oplus$ 0)&$\IZdos (\oplus$ 0)&$\IZ (\oplus$ 0)&0&0&0&$\IZ$\\
\hline
(-1)&0&$\IZ$&0&$\IZdos (\oplus$ 0)&$\IZdos (\oplus$ 0)&$\IZ$&0&0&0\\ \hline\hline
\end{tabular}
\end{center}
\end{table}


\section{(Co)homology and D-branes in Orientifolds}

Before interpreting physically the RR fields shown in tables \ref{Kop-} and \ref{Kop+} we must answer
the second question raised in the previous section: What is the cohomology groups which classify RR
fields in the presence of $d$-branes and $Op-$planes (for $d\leq p$)?

In this section we give the answer by wrapping D$(d+n)$-branes on homology $n$-cycles. To obtain the
cohomology groups we first find their associated homology groups and by Poincar\'e duality we can
find them. Our aim is to compare these results to those obtained by K-theory in the
previous section by using the AHSS. This will be the goal of the next section.

\subsection{Wrapping D-branes on homological cycles}

Wrapping D$(p+n)$-branes in non-trivial and compact homology $n$-cycles of projective
spaces\footnote{In fact, they are not truly homological cycles in the bulk space to the orientifold, 
unless we
are removing the origin. In this construction, we remove the origin in order to obtain stable branes
by wrapping them on non-trivial homological cycles \cite{mms}.} has been used extensively to
classify fluxes which give rise to different kind of orientifold planes \cite{hananyo,bgs,sugimoto}. 

As we have seen in previous sections, new types of orientifold planes (\RROp) appear when discrete RR
fluxes are turned on. These fluxes are classified by the cohomology of projective spaces (the
transversal spaces to the orientifolds), i.e., by the group $H^{6-p}(\IRP^{8-p};\IZ (\ItZ))$. The
``brane realization of RR discrete torsion'' is obtained by intersecting a D$(p+2)$-brane and an
$Op^\pm$-plane at one point. Then, it is possible to deform the D-brane in such a way that it wraps 
on a
two-cycle of $\IRP^{8-p}$. If the origin is not removed, the two-cycle is not a truly homological
cycle of the bulk space and it shrinks to a point, giving rise to our original configuration of a
D$(p+2)$-brane intersecting the $Op^\pm$-plane. However, if the origin is removed, we actually are
allowed to wrap D$(p+2)$-branes on homological two-cycles of $\IRP^{8-p}$ to get an \RROp
-plane. Moreover, according to Ref. \cite{sugimoto}, a D$(p+2)$-brane wrapped on a two-cycle carries a
RR
charge of a D$p$-brane. So, what we have indeed, is that after the wrapping process, we get a truly
D$p$-brane where the RR charge is given directly by the value of the (co)homology group which 
classifies
the two-cycle wrapped by the D$(p+2)$-brane.

This is indeed the idea we want to use in order to obtain the cohomology groups which classify RR
fields in the presence of orientifold planes and lower dimensional branes on top of them.

\subsubsection{$Op$-planes and $p$-branes}
Let us
start by {\it re-obtaining} the cohomology groups which classify orientifolds. According to the
``brane realization picture of discrete RR fluxes", we must take a D$(p+2)$-brane and wrap it on a
two-cycle of $\IRP^{8-p}$ (notice that this cycle can be twisted or untwisted). However, it turns out
that the non-zero valued homological group classifying two-cycles is
$H_2(\IRP^{8-p};\ItZ)=\IZdos$, which actually classifies {\it twisted} cycles. 
The fact that we require twisted cycles, can also be understood from
a physical perspective:  D$(p+2)$-branes couple with $(p+3)$-forms in the bulk space, and
according to relations (\ref{un-twisted}), these forms are in fact, {\it twisted}. So, a
D$(p+2)$-brane can only be wrapped on twisted cycles. Finally, by using Poincar\'e duality, which reads,
\beqa 
\begin{array}{cl} 
\hbox{For $n$ odd:}&H_i(\IRP^n; \IZ (\ItZ))\cong H^{n-i}(\IRP^n;\IZ(\ItZ)),\\ 
\hbox{For $n$ even:}&H_i(\IRP^n; \IZ (\ItZ))\cong
H^{n-i}(\IRP^n;\ItZ(\IZ)), 
\end{array} 
\eeqa
we find that the cohomology group which classifies orientifold planes (when discrete RR fluxes are
turned on) is actually $H^{6-p}(\IRP^{8-p};\ItZ)=\IZdos$. Notice that, according to this procedure, the
above cohomology group is also {\it the one which classifies RR fields in the presence of
$Op^\pm$-planes and $p$-branes}. Let us fix the notation: a $p$-brane stands for a generic
$p$-dimensional brane, while one with discrete $\IZdos$ topological charge, will be denoted as a
$\widehat{p}$-brane. This notation stands from the fact that up to this point we do not the nature of
these
objects. We require K-theory in order to get a more precise description of them.

Thus, we can get (as a first approximation) a picture of an \RROp-plane as one $Op^\pm$-plane plus a
$\widehat{p}$-brane. Of course, this turns out to be not correct at all, since an \RNOp-plane is given
by an $Op^-$ plus a half stuck brane, $\frac{1}{2}Dp$. Notice also that this description is valid just
for the case $2\leq 8-p$, i.e., for $p\leq 6$.

There is a second possibility to get a $p$-brane (or a $\widehat{p}$-brane) by wrapping D-branes on
homology cycles. This is given by wrapping a D$(p+6)$-brane on a 6-cycle of $\IRP^{8-p}$. The homology
group classifying such cycles is $H_6(\IRP^{8-p})=\IZdos$. By Poincar\'e duality this is the cohomology
group $H^{2-p}(\IRP^{8-p})=\IZdos$ which actually classifies other type of orientifold planes denoted
as $\widehat{Op^\pm}$. Notice that this is possible just for the case $0\leq 2-p$. i.e., for $p\leq 2$.
By the same argument as above, the exotic orientifold plane $\widehat{Op^\pm}$ can be expressed as the
sum of an $Op^\pm$-plane plus a $\widehat{p}$-brane and moreover, for $p\leq 2$ we actually have 8
different types of orientifold planes by taking all the possible combination of RR discrete fluxes.
Hence, our description of wrapping branes on homology cycles reproduce all these well-known results.
Our goal, for the next section is to describe the nature of the $\widehat{p}$-branes and establish a
difference between the ones associated to \RROp and $\widehat{Op^\pm}$.

From the above analysis we get two important results: 1) we have a procedure to classify by cohomology
all the spectrum of RR fields in the presence of orientifolds $Op^\pm$ and $d$-branes, and 2) we
require to classify them by K-theory in order to {\it refine} our conclusion of what an \RROp-plane or
an $\widehat{Op^\pm}-plane$ are made of.

\subsubsection{$Op$-planes and $d$-branes}
Let us start by working out the point 1). Our goal is to extend this idea to any $d$-brane on top of an
$Op$-plane, with $d<p$. This means that we will be able (in the cohomology sense) to obtain $d$-branes
by wrapping D$(d+n)$-branes on non-trivial compact homological $n$-cycles\footnote{We are restricting
ourselves to the study of D-branes completely immersed in the orientifold plane.}. In order to do that,
we require to know what homological cycles are suitable for wrapping D-branes on them, as was done for
the $O6$-plane in Ref. \cite{yo}. The answer is given by the relations (\ref{un-twisted}).

Far away from the orientifold plane and locally, the relevant string theory is the Type II one (A or
B depending of the dimension of the orientifold plane). The RR forms $C_{d+n+1}$ couple to D$(d+n)$-branes,
and they are affected by the orientifold projection as in equations (\ref{un-twisted}). According to
the nature of the RR form, twisted or untwisted, the associated brane can be wrapped on a homological
cycle of the same nature, i.e., a brane which couples to a (un)twisted form, can be wrapped only on a
(un)twisted cycle. The RR D$d$-brane charge of a D$(d+n)$-brane wrapped on a non-trivial homological
$n$-cycle is the same that the corresponding $n$-th homology group value of $\IRP^{8-p}$.  Finally,
by Poincar\'e duality, we can obtain the relevant cohomology group for such $d$-branes.\\

\noindent
{\bf An example: The cohomology of $O5$}.

In order to give an specific example, take for instance the $O5$-plane. By Eq. (2.7) we know that D7,
D3 and D(-1)-branes for Type IIB theory can be wrapped only on twisted cycles and D9, D5 and D1-branes on
untwisted ones. Then, the homology groups of $\IRPthree$ are given by
\beqa
\begin{array}{cc}
H_0(\IRPthree , \IZ)\, =\, \IZ, \quad & H_0(\IRPthree , \widetilde{\IZ})\,=\, \IZdos,\\
H_1(\IRPthree , \IZ)\,=\,\IZdos, \quad & H_2(\IRPthree , \IZ )\,=\,\IZdos,\\
H_3(\IRPthree , \IZ)\,=\,\IZ. \quad &\\
\end{array}\nonumber
\eeqa
Now, by wrapping
D$(d+n)$-branes (with $0\le n \ne 3$) we obtain the desired $d$-branes. For instance,
wrapping D3 and D(-1)-branes on
the twisted 0-cycle we obtain states that are identified with $\widehat{3}$ -  and $\widehat{(-1)}$
-branes
(since the 0-cycle has $\IZdos$-charge). If now we wrap  D7 and D3 on  twisted
two-cycles we get $\widehat{5}$ and $\widehat{1}$ -branes. On the other hand, wrapping D5 and D1 branes
on untwisted 0, 1 and 3-cycles, we obtain 5 and 1, $\widehat{4}$ and $\widehat{(-1)}$, and 2 branes
respectively.

For completeness and future reference, we proceed similarly for all orientifolds $Op$ with 
$p \le 6$. The results are listed in tables \ref{coh1} and \ref{coh2}.

\begin{table}
\begin{center}
\caption{The table shows the twisted and untwisted $n$-cycles in where suitable D$(d+n)$-branes can 
be wrapped.}
\label{coh1}     

\begin{tabular}{||c|c|c||}\hline\hline
$Op$-plane&On untwisted cycles&On twisted cycles\\\hline\hline
$p=6,2$&D6,D2&D8,D4,D0\\\hline
$p=5,1$&D9,D5,D1&D7,D3,D(-1)\\\hline
$p=4,0$&D8,D4,D0&D6,D2\\\hline
$p=3$&D7,D3,D(-1)&D9,D5,D1\\\hline\hline 
\end{tabular}
\end{center}
\end{table}

\begin{table}
\begin{center}
\caption{D$d$-branes obtained by wrapping D$(d+n)$-branes on $n$-cycles. We label as D$d$-branes the branes which are also classified by
K-theory. The other ones are labeled just as $d$-branes. }
\label{coh2}

\begin{tabular}{||c||c|c||c|c||}\hline\hline
$Op$-planes&$H_n(\IRP^{8-p};\IZ)$&D$d$-branes&$H_n(\IRP^{8-p};\ItZ)$&D$d$-branes
\\\hline\hline
6&$H_0(\IRP^{2};\IZ)=\IZ$&D6\quad D2&$H_0(\IRP^{2};\ItZ)=\IZdos$&\Dfour\quad\Dzero\\
 &$H_1(\IRP^{2};\IZ)=\IZdos$&\Dfive\quad\Done&$H_2(\IRP^{2};\ItZ)=\IZ$&D6\quad
D2\\\hline
5 &$H_0(\IRP^{3};\IZ)=\IZ$&D5\quad
D1&$H_0(\IRP^{3};\ItZ)=\IZdos$&\Dthree\quad\Dmone\\
 &$H_1(\IRP^{3};\IZ)=\IZdos$&\Dfour\quad\Dzero&$H_2(\IRP^{3};\ItZ)=\IZdos$&
$\widehat{1}$\quad $\widehat{5}$\\
 &$H_3(\IRP^{3};\IZ)=\IZ$&2&&\\\hline
4&$H_0(\IRP^{4};\IZ)=\IZ$&D4\quad D0&$H_0(\IRP^{4};\ItZ)=\IZdos$&\Dtwo\\
 &$H_1(\IRP^{4};\IZ)=\IZdos$&\Dthree\quad\Dmone&$H_2(\IRP^{4};\ItZ)=\IZdos$&
$\widehat{4}$\quad$\widehat{0}$\\
 &$H_3(\IRP^{4};\IZ)=\IZdos$&$\widehat{1}$
 &$H_4(\IRP^{4};\ItZ)=\IZ$&2\\\hline
3&$H_0(\IRP^{5};\IZ)=\IZ$&D3\quad D(-1)&$H_0(\IRP^{5};\ItZ)=\IZdos$&\Done\\
 &$H_1(\IRP^{5};\IZ)=\IZdos$&\Dtwo&$H_2(\IRP^{5};\ItZ)=\IZdos$&$\widehat{3}$\\
 &$H_3(\IRP^{5};\IZ)=\IZdos$&$\widehat{0}$&$H_4(\IRP^{5};\ItZ)=\IZdos$&$\widehat{1}$\\
 &$H_5(\IRP^{5};\IZ)=\IZ$&2&&\\\hline
2&$H_0(\IRP^{6};\IZ)=\IZ$&D2&$H_0(\IRP^{6};\ItZ)=\IZdos$&\Dzero\\
 &$H_1(\IRP^{6};\IZ)=\IZdos$&\Done&$H_2(\IRP^{6};\ItZ)=\IZdos$&$\widehat{2}$\\
 &$H_3(\IRP^{6};\IZ)=\IZdos$&$\widehat{(-1)}$&$H_4(\IRP^{6};\ItZ)=\IZdos$&$\widehat{0}$\\
 &$H_5(\IRP^{6};\IZ)=\IZdos$&$\widehat{1}$&$H_6(\IRP^{6};\ItZ)=\IZ$&D2\\\hline
1&$H_0(\IRP^{7};\IZ)=\IZ$&D1&$H_0(\IRP^{7};\ItZ)=\IZdos$&\Dmone\\
 &$H_1(\IRP^{7};\IZ)=\IZdos$&\Dzero&$H_2(\IRP^{7};\ItZ)=\IZdos$&$\widehat{1}$\\
 &$H_5(\IRP^{7};\IZ)=\IZdos$&$\widehat{0}$&$H_4(\IRP^{7};\ItZ)=\IZdos$&$\widehat{(-1)}$\\
 &&&$H_6(\IRP^{7};\ItZ)=\IZdos$&$\widehat{1}$\\\hline
0&$H_0(\IRP^{8};\IZ)=\IZ$&D0&$H_2(\IRP^{8};\ItZ)=\IZdos$&$\widehat{0}$\\
&$H_1(\IRP^{8};\IZ)=\IZdos$&\Dmone&$H_6(\IRP^{8};\ItZ)=\IZdos$&$\widehat{0}$\\
&$H_5(\IRP^{8};\IZ)=\IZdos$&$\widehat{(-1)}$&&\\\hline\hline
\end{tabular}
\end{center}
\end{table}

As we can see from these tables, in general there are three different types of cohomology groups classifying RR forms in the presence of
$Op$-planes and $d$-branes. This is as follows:
\begin{itemize}
\item
$H^{8-p}(\IRP^{8-p})=\IZ$. This group classify RR forms related to D$p$-branes on top of $Op^\pm$-planes. Give us the usual integer RR charge
of such branes.
\item
$H^{6-d}(\IRP^{8-p})=\IZdos$. It classifies RR forms related to $\widehat{d}$-branes on top of
$Op$-planes. Notice that in the case $d=p$, the
non-trivial class of $\IZdos$ stands for the presence of an \RROp -plane. This is true for $p-d\leq 2$ and $d\leq 6$.
\item
$H^{2-d}(\IRP^{8-p})=\IZdos$ ($d\neq 2$). It also classifies RR forms related to $\widehat{d}$-branes.
Notice that in the case $d=p$, we recover
the classification of $\widehat{Op^\pm}$-planes. This case is valid just for $d<2$ and $p-d\leq 6$. In the case of $d=2$ the cohomology value
is integer, and it is related to $d$-branes.
\end{itemize}

This in turn show us that we have actually two different cohomology groups for a $\widehat{d}$-brane
given by $H^{6-d}$ and $H^{2-d}$ in
the case $0<p-d\leq 2$. However we also have a single cohomology group pointing out the presence of two different branes. $H^{6-d}$ is
related to $\widehat{d}$-branes as well as to $\widehat{(d\pm 4)}$-branes, since $H^{6-d}=H^{2-(d-4)}$
(or vice versa, $H^{2-d}=H^{6-(d+4)}$).

In order to get a more exact picture, let us take advantage of our knowledge of K-theory. The T-dual versions of D-branes in Type I and Type
$Usp(32)$ theories (recall, just those with no transversal coordinates to $Op$) are given as follows:
\begin{itemize}
\item
In the presence of an $Op^-$-plane, we actually have D$p$, $\widehat{{\rm D}(p-1)}$, $\widehat{{\rm D}(p-2)}$ and D$(p-4)$ -branes (since we
are considering just orientifolds which dimension is less than seven).
\item
In the presence of an $Op^+$-plane, we have $\overline{{\rm D}p}$, D$(p-4)$, $\widehat{{\rm D}(p-5)}$ and $\widehat{{\rm D}(p-6)}$ -branes.
\end{itemize}

Hence we conclude that some of the branes given in table \ref{coh2} are in fact the D-branes contained in the above spectrum of branes. In
particular,
\begin{itemize}
\item
$\widehat{d}$-branes classified by $H^{6-d}$ are in fact $\widehat{{\rm D}d}$-branes on top of an
$Op^-$-plane.
\item
$\widehat{d-4}$-branes classified by $H^{2-(d-4)}$ are in fact $\widehat{{\rm D}(d-4)}$-branes on top of an $Op^+$-plane.
\end{itemize}

Finally, notice that topologically is allowed to relate $d$-branes with $Op^+$-planes and $(d-4)$-branes with $Op^-$-planes. We will discuss
this possibility in the next section.

\subsection{R-R and NS-NS fluxes}

In order to prove that some of the branes obtained by wrapping higher or equal dimensional branes on
suitable non-trivial homological cycles, are truly the T-dual version of the known D-branes
classified by K-theory in Type I and $USp(32)$ string theories, we will use the topological relation
between products of RR and NS-NS fluxes in Type II theories and D-branes, studied in \cite{urangaf}.

Let us describe briefly the procedure which transforms
topologically a non-BPS $\widehat{{\rm D}d}$-brane into a source-flux given by $H_{NS}G_{6-d}$ for Type II theories.

For Type II theories these couplings are given by
\beqa
\int_{{\cal M}_{10}} H_{NS}G_{6-d}C_{d+1},
\label{HGC}
\eeqa
with $G_{6-d}$ being the RR field strength of $C_{5-d}$ (with appropriate $d$ for IIA or IIB
theories). Topological couplings given by (\ref{HGC}) show that there is the possibility to endow
NS-NS and RR fluxes with charges under RR fields $C_{d+1}$, justly as D$d$-branes. Thus,
transitions between branes and configurations of suitable fluxes are possible. In \cite{urangaf}
the case of the non-BPS \Dfour -brane was considered. Since these two systems are topologically equivalent, 
we are able to invert the procedure, i.e.,
having a source-flux of the form $H_{NS}G_{6-d}$, we can transform it into a D$d$-brane of Type II
theory.

For T-dual versions of branes on top of an $O9^+$-plane, we consider the product of fluxes, far
away from the orientifold, $H_{(7)}G_{2-d'}$, with $H_{(7)}$ being the magnetic dual of $H_{NS}$ and $d'=d-4$.
This is because in the presence of an $Op^+$-plane there is a magnetic NS-NS field in the
background (remember that an $Op^+$-plane is constructed by a NS5-brane intersecting an orientifold
plane $Op^-$). The product $H_7 G_{6-d'}$ can topologically be transformed into a $\widehat{{\rm D}d'}$-brane.

By this procedure we are able to confirm our proposal concerning that cohomology groups (Poincar\'e 
dual of those given in
table \ref{coh2}), are the relevant ones for D$d$-branes on top of $Op$-planes.

For more details concerning the characteristics of these branes, see appendix A.


\section{Physical Interpretation of RR Fields in K-theory}

Up to here we have classified all RR-fields in a background given by $Op^\pm-$planes and $d$-branes by
using K-theory. Also we have classified RR fields in the presence of $d$-branes on top of orientifold
planes through cohomology.  In this section we relate both descriptions by using the AHSS which in turn
provides a physical interpretation of such $d$-dimensional subspaces when the fluxes are turned on.

\subsection{$d$-branes as D$d$-branes}

The K-theory classification of RR fields given in tables \ref{Kop-} and \ref{Kop+} give us a lot of
information. For instance, as it was said, for the case $d=p$ we get a truly classification of
orientifold planes. In a different (but related) point of view, such a classification give us the
possible $p$-branes present on top of $Op$-planes when discrete RR fluxes are turned on. By
considering this latter alternative description, we can interpret the exotic orientifold planes \RROp
as composed by a ``normal" $Op$-plane and a $p$-brane with certain RR charge (integer or discrete).
However, for the case of $d<p$, the interpretation is not so obvious as the above one. Firstly, we
notice that the K-theory classification of RR fields given in tables \ref{Kop-} and \ref{Kop+} is
different from the RR charge K-theory classification. In fact, the discrepancy is given by the second
terms in the right-handed side of the fields in the above tables. Recall that these fields came from
the second K-theory groups in the rhs in equations (\ref{formula}). On the other hand, the lhs terms in
tables \ref{Kop-} and \ref{Kop+} which came from the first terms in the rhs of equations
(\ref{formula}), actually give us the RR charges of {\it Dd-branes} on top of $Op$-planes. This can be
easily inferred by noticing that such K-theory groups give the same result than the K-theory groups
(classifying RR charges) given in eqs. (\ref{cargas1})  and (\ref{cargas2}).

Hence, in this
case we can interpret physically the meaning of the $d$-dimensional submanifold related to the RR
fields classified by K-theory. They are justly the D$d$-branes with RR charge, i.e., the RR fields
computed by $KR^{d-p}(\{pt\})$ for $Op^-$ and $KH^{d-p}(\{pt\})$ for $Op^+$ (see equation
(\ref{formula})) have D$d$-branes as sources. Then, our conclusion reads:

{\it $d$-branes associated to RR fields classified by K-theory through $KR^{d-p}(\{pt\})$ for
$Op^-$ and $KH^{d-p}(\{pt\})$ for $Op^+$, are exactly the usual D$d$-branes on top of orientifold
planes.  They are the T-dual version of the D-branes on top of $O9^-$ and $O9^+$ planes classified by
Eqs. (\ref{cargas1}) and (\ref{cargas2}).}\footnote{As it was said, T-duality is taken on longitudinal coordinates on the D-branes on
the ten-dimensional theories, Type I and Type $USp(32)$. When the number of compact coordinates is higher than the dimensionality of the
D-brane, we get a brane which has some transversal coordinates to the orientifold plane (the T-dual version of $O9$). Such branes are not
considered in this paper.}

\subsection{$d$-branes as $d$-fluxbranes?}

Lets turn our attention to the RR fields not associated with a source. So, it is time to answer our
third question raised in section 4.3. For that, let us start by giving a ``cohomology" approach of the
answer. We saw that RR fluxes $G_{6-d}$ which are classified by the cohomology group $H^{6-d}$ are
related to truly $\widehat{{\rm D}d}$-branes on top of $Op^-$-planes. This conclusion was taken after
using our K-theory knowledge of RR classification and by topologically transforming the product of
fluxes $G_{6-d}H_{NS}$ into a $\widehat{{\rm D}d}$-brane.

As we said at the end of section 5, we can relate the fluxes $G_{6-d}$ to $(d-4)$-branes not in the
presence of an $Op^+$ but in the presence of an $Op^-$ instead. Clearly, this $(d-4)$-brane can not be
a Dirichlet brane, since we know by K-theory which branes are present in top of an orientifold plane.
Hence, whatever these branes could be, they are associated to RR fields without source. Such RR fields
are constructed far away from the orientifold plane, by the product of fluxes $G_{2-d'}H_{(7)}$, where
as usual, $d'=d-4$. Notice that in this situation, although we have an $Op^-$-plane, we take the
product of the RR flux with the magnetic dual of $H_{NS}$. After all, this product is topologically
available. Physically can be understood as the presence of a NS5-brane far away from the orientifold.

On the other hand,  RR fields that are not associated to any source (i.e., without any 
$d'$-dimensional objects charged under this field) can be only tangent to the
$d'$-brane. This topological property allows to avoid sources for the
fields.  According to \cite{moore-witten}, this tangent field denoted as $F_{9-d'}$
must satisfy that
\beqa
\int_{\partial{\cal M}_{9-d'}} F_{9-d'} \; <\;\infty\;,
\eeqa
with ${\cal M}_{9-d'}$ the $(9-d')$-dimensional transverse space to a $d'$-dimensional
object.
If this field is extended over ${\cal M}_{9-d'}$ it does not require sources. 

Now, in order to fix the notation, let us classify RR fields (by cohomology), with and without source, related to the same $d$-dimensional
brane. Such RR fields (as we said, there are actually two cohomology groups classifying RR fields related to a $d$-brane) are classified by
$H^{6-d}$ and $H^{2-d}$. The latter one refers to RR fields that do not have a source in the presence of an $Op^-$-plane. The opposite
situation holds for an $Op^+$.

We argue that, this is the case for the RR fields given by the groups $KR^{d-10}(pt)$ and
$KH^{d-10}(pt)$ in equation (\ref{formula}) (or, for the right handed fields in tables \ref{Kop-} and
\ref{Kop+}). 

Hence, because they are source-free, they can be extended over ${\cal M}_d$ and
therefore,

\beqa
\int_{{\cal M}_{9-d}}F_{9-d} \; < \; \infty.
\eeqa
This is precisely the property that a fluxbrane satisfies.

A flux $d$-brane (see \cite{fluxbranes,costa,urangawf,fluxandbranes}), denoted as F$d$-brane, is a
$(d+1)$-dimensional object with non-zero flux $F_{9-d}$ on the 
$(9-d)$-dimensional transverse space to the brane. This is contrasted with the usual
D$d$-branes which carry a RR charge measured by integrating out the field
strength over a surrounding sphere. Also, fluxbranes are generalizations in
higher dimensions of flux-tubes, that are solutions in General Relativity with
precisely these properties. The most known example of it is the Melvin universe
\cite{melvin}. Basically this consists in a solution of the Einstein's equation
for General Relativity in four dimensions, in where a 2-form field is present in
the background and it is confined by its self-gravity.

{\it We argue that the RR fields classified by $KR^{d-10}(\{pt\})$ for $Op^-$ and $KH^{d-10}(\{pt\})$ 
for
$Op^+$( see equation (\ref{formula}))  are actually the field strength $F_{9-d}$ related to
fluxbranes. This is, the role of the $d$-dimensional subspaces for this kind of RR fields without
source, is the presence of a flux d-brane $Fd$, or F$d$-brane for short.}

The interesting fact is that cohomology also captures the presence of the flux $F_{9-d}$ by some
unknown mechanism. Some of the objects classified by cohomology seem to be D-branes at that level, but
in K-theory are related to source-free RR fields. A more deeper study of this features is required but
it is beyond the scope of this paper.

From now on, we will denote the $d$-dimensional objects related to RR fields without source as
``F$d$''-branes which has related a $\IZdos$ field. This notation remarks our limited knowledge about
their nature.

\subsection{Example. Branes and Fluxes in the $O5$-plane}

We are ready to apply all the information we have got in the previous sections. On one hand we have the
cohomological classification of RR fields in the presence of $Op$-planes and $d$-branes. Also, we were
able to infer some of the properties of such branes and the role they are playing on. The same was done
in the case of the K-theory classification of RR fields given in section 4. The final step is to relate
both of them by the AHSS as was done in Ref. \cite{bgs} by BGS.

Let us do it by analyzing a concrete example: the orientifold five-plane. We will analyze the case for
each value of $d$ in the presence of an $O5$-plane. The case $d=5$ has already been studied in previous
sections, although there is some extra information which is important to point out.

\noindent
{\bf Five brane}

\noindent
According to our discussion at the beginning of this section, it is possible to describe the exotic orientifold five-planes as:

\beqa
\begin{array}{cl}
\hbox{\RNOfive}=&O5^- +\frac{1}{2}\rm{D}5,\\
\hbox{\RPOfive}=&O5^+ +\hbox{``F5''}.
\end{array}
\eeqa

We do not know exactly what ``F5'' could be, but as it is classified by the second term in the left
hand side of (\ref{formula}), which corresponds to a RR field without a source. We argue that this is a
fluxbrane F5 with $\IZdos$ charge and moreover, obeys a T-duality relation given by Eq.
(\ref{Tnobranas}) (see appendix B), at the cohomology level.

\vskip .5truecm

\noindent
{\bf Four-brane}

\noindent
According to table \ref{coh2}, the (co)homology group for a 4-brane on an
$O5$-plane is given by
\beq
H_1(\IRPthree , \IZ ) \cong H^2(\IRPthree , \IZ)=\IZdos \, .
\eeq
The K-theory groups are
\beqa
\begin{array}{cl}
O5^-:&KR^{-6}(\IS^{4,0})=\IZdos,\\
O5^+:&KH^{-6}(\IS^{4,0})=\IZdos.
\end{array}
\eeqa
Now, we can proceed to built the corresponding sequence in order to
resolve the extension problem addressed by the AHSS. $d_3$ is also trivial for both cases, and we find that
\beqa
\begin{array}{rcl}
K_0=K_1=K_2&=&\IZdos\quad\hbox{for both cases}\\
K_2/K_3&=&H^2=\IZdos\\
K_3&=&0\;.
\end{array}
\eeqa
The extension problem is given by the exact sequence
\beqa
\diagram{
0&\harr{}{}&K_3&\harr{}{}&K_2&\harr{}{}&K_2/K_3&\harr{}{}&0\cr
&&\parallel&&\parallel&&\parallel&&\cr
&&0&&\IZdos&&\IZdos&&\;.
}
\eeqa
This is trivial and it is concluded that there are not effects on both $O5^{\pm}-$planes, due to the
torsion flux $G_2$, i.e, cohomology and K-theory descriptions coincide. For the $O5^-$-plane, this is
the T-dual version of the \Deight -brane in Type I theory, while for the $O5^+$-plane, the presence
of a topological 4-dimensional object is unexpected. As for the five branes, we can interpret this
brane as the result of turning on a discrete RR field (without sources) over a 4-dimensional
submanifold of the orientifold five-plane. We argue that this is related to a 4-fluxbrane (or a
``F4''-brane). It would be very interesting the study of anomalies in these objects and their 
relation to
anomalies of fluxes described in \cite{urangaf}. According to equation (A.2) in the appendix A, this
4-brane is T-dual related to a 4-brane on an $O4^+$- and $O6^+$-planes; this is obtained by the
Eq. (\ref{Tnobranas}) at the cohomology level.

\vskip .5truecm
\noindent
{\bf Three-brane}\\
The cohomology group which classifies three branes on top of $O5$-planes is $H^3(\IRPthree, \IZ)$, 
and the K-theory groups are given by
\beqa
\begin{array}{cl}
O5^-:&KR^{-7}(\IS^{4,0})=\IZdos,\\
O5^+:&KH^{-7}(\IS^{4,0})=0.\\
\end{array}
\eeqa
In the  case of an $O5^+$-plane, the map $d_3:
H^0(\IRPthree)\rightarrow\widetilde{H}^3(\IRPthree)$ is surjective;  this means that the flux $G_3$
is lifted to a trivial class in K-theory. Physically this means that there are not any type of three-branes on top of an $O5^+$-plane
(neither D-branes nor ``fluxbranes"). For the $O5^-$, $d_3$ is trivial and the
extension problem is given by
\beqa
\diagram{
0&\harr{}{}&K_4&\harr{id}{}&K_3&\harr{id}{}&K_3/K_4&\harr{}{}&0\cr
 &         &\parallel&     &\parallel&     &\parallel&       &\cr
 &         &0&             &KR^{-7}(\IS^{4,0})=\IZdos&        &H^3(\IRP^3;\ItZ )=\IZdos&          &\cr
}
\eeqa
The extension is trivial and we conclude that this brane is the T-dual version of the \Dseven -brane in Type I theory.

\vskip .5truecm
\noindent
{\bf Two-brane}\\
Possible two-branes are obtained by wrapping a D5-brane on the non trivial
untwisted and compact 3-cycle of $\IRPthree$. The 3-cycle is classified by the
untwisted homology group
\beq
H_3(\IRPthree , \IZ) \cong H^0(\IRPthree , \IZ)= \IZ.
\eeq
However this integral flux has another interesting interpretation. As was pointed 
out in \cite{bgs, sugimoto}, this flux is related to massive IIA supergravity
\cite{sugram,sugram2}.

In order to look for some correlations, we need to solve the
extension problem given by the AHSS. In the case of $O5^-$ this reads,
\beqa
\diagram{
0&\harr{}{}&K_1&\harr{}{}&K_0&\harr{id}{}&K_0/K_1&\harr{}{}&0\cr
 &         &\parallel&   &\parallel&     &\parallel&       &\cr
 &         &0&           &KR^{-8}(\IS^{4,0})=\IZ&           &H^0(\IRP^3;\IZ)=\IZ&&\cr
}\:.
\eeqa
This is trivial and admits just one solution (the trivial one). The integer flux described by K-theory
indicates the presence of massive D2-branes \cite{massivebranes, massivebranes2}. Moreover, for the
$O5^+$-plane there is a surjective map $d_3:H^0=\IZ \rightarrow \widetilde{H}^3=\IZdos$ which implies
that odd values of $G_0$ are not allowed.  This must be related to an anomaly in the three-dimensional
gauge theory on 2-branes on top of an $O5^+$-plane with odd $G_0$.  These two-branes could be related
to two-fluxbranes. It would be very interesting to study these systems and their possible anomalies.

\vskip .5truecm
\noindent
{\bf One-brane}\\
Essentially we have the same cohomology and K-theory groups as for the five-branes on both kind of
orientifolds. However the difference is that the K-theory groups are inverted respect to
the five-branes. We are not describing our calculations in detail but just focusing in the results 
and in their physical interpretation.

For the $O5^-$-plane we have a D1-brane (the usual one) carrying an integer RR
charge. Also we have an induced ``F1''-brane.  For the $O5^+$-plane we have also
the usual $D1$-brane expected by T-duality, that corresponds to the D5-brane on Type
$USp(32)$ string  theory, and a fractional integer one-brane, $\frac{1}{2}$D1-brane.

\vskip .5truecm
\noindent
{\bf Zero-brane}\\
In this case we have the same situation as in the case for the 4-branes. The result is that for the
$O5^-$-plane we have an induced ``F0''-brane with topological charge $\IZdos$. For the $O5^+$-plane
we have the expected \Dzero-brane. ``F0''-brane obeys a T-duality relation given by Eq.
(\ref{Tnobranas}).

\vskip .5truecm
\noindent
{\bf (-1)-brane}\\

The case of the (-1)-brane is very interesting and we analyze it in more detail. According to table
\ref{coh2}, the cohomology group which classifies RR fields related to $(-1)$-branes is
$H^3(\IRP^3,\ItZ )=\IZdos$.

For the
$O5^+$-plane,
there exists a surjective map 
\beqa
d_3 : H^0(\IRP^3,\IZ )= \IZ \rightarrow H^3(\IRP^3,\ItZ )=\IZdos
\eeqa
and the flux $G_3$ is lifted to a trivial element in K-theory. This means that a $(-1)$-brane is
unstable and decay to vacuum when an $O5^+$ is present. However, in order to get a better picture of
this situation, we must resolve the extension problem addressed by the AHSS. We found that
$K_3=KH^{-11}(\IS^{4,0})=\IZdos$ and that $K_4=K_3/K_4=0$. Hence, the sequence would be trivial only if
the K-theory class was zero. Physically this means that although K-theory actually classifies the RR
charge of $(-1)$-branes, it also establishes an extra condition: only an even number of $(-1)$-branes
is allowed to be on top of an $O5^+$-plane, i.e., (since the RR charge of such branes is $\IZdos$) the
K-theory charge must be cancelled.

This resembles the behavior of the \Dthree-brane in $USp(32)$ theory where a single three-brane is
unstable, but it cannot decay to the vacuum because it has a discrete $\IZdos$ charge. Thus, it is
expected that K-theory measures this charge, but it does not allow the presence of a single non-BPS
D-brane.  This is actually the required topological condition on the \Dthree -brane on top of an
$O9^+$-plane placed at a point in $\IT^6$, described in section 3, in order to cancel global gauge
anomalies on suitable probe branes. Here we have the same condition applied to a T-dual version of such
a system (notice that by considering a T-dual version of $USp(32)$ string theory, we are actually
compactifying the system on a torus $\IT^{4}$ where the $(-1)$-brane is placed at a point). As was
shown in Refs. \cite{yo, bergmanyo}, this is also a property of the \Dseven-brane in Type I theory.

The case of the $Op^-$ is puzzled. We obtain a non-zero value by cohomology but it is zero by
K-theory. As K-theory is given exactly by the graded complex and then by cohomology, this is somewhat
contradictory. We do not know how to explain this feature, although we think that a more deeper study
on differences at the cohomology level for branes on top of $Op^-$ or $Op^+$, could be very helpful
in order to explain the above puzzle. Notice however that $(-1)$-branes given by cohomology actually
reproduce the expected $(-1)$-branes classified by K-theory.

Finally, for all different values of $d$ and $p$, we resume our results in table \ref{todas}.

\begin{table}
\begin{center}
\caption{Brane states on top of $Op^-$-planes considering discrete RR
fields. {\small Left superscript $m$ stands for the massive
D2-branes. In the
case of D6, this is fractional and for the $O6^+$ the flux is twice an integer
\cite{bgs}. For the $Op^+$-planes, D$p$-branes are actually
anti-D$p$-branes by tadpole cancellation in $USp(32)$ string theory.}} 
\label{todas}
\begin{tabular}{||c|c|c|c||}\hline\hline
&&&\\
$O6^-$&$O5^-$&$O4^-$&$O3^-$\\\hline
D6+$^m$D6&&&\\
\Dfive&D5+$\frac{1}{2}$D5&&\\

\Dfour&\Dfour&D4+$\frac{1}{2}$D4&\\
-&\Dthree&\Dthree&D3+$\frac{1}{2}$D3\\
D2+$^m$D2&$^m$D2&\Dtwo +$^m$D2&\Dtwo +$^m$D2\\
``F1''&D1+``F1''&``F1''&\Done +``F1''\\

``F0''&``F0''&``F0'' + D0&``F0''\\\hline\hline\hline
&&&\\
$O6^+$&$O5^+$&$O4^+$&$O3^+$\\\hline
D6+$^m$D6&&&\\
``F5''&D5+``F5''&&\\
``F4''&``F4''&D4+``F4''&\\
-&-&-&D3\\
D2+$^m$D2&$^m$D2&$^m$D2&$^m$D2\\
\Done&D1+ $\frac{1}{2}$D1&-&-\\
\Dzero&\Dzero&D0 + $\frac{1}{2}$D0&-\\
-&\Dmone&\Dmone&D(-1)\\\hline\hline
\end{tabular}
\end{center}
\end{table}

\subsection{The Case for $p\le 2$}
It was shown in \cite{hananyo,bgs} that for $p\le 2$ there are some extra interesting
features for
both orientifolds $Op^\pm$.
In this case there are additional RR discrete fluxes classified by the cohomology group
$H^{2-p}$. Let us summarize some results given in \cite{bgs}:

\begin{itemize}

\item
In the case of $O2^-$ we have actually three fluxes to be considered. $G_6$ stands for the presence of
a BPS D2-brane on top of the orientifold. $G_4$ is the one related to the exotic plane \RNOtwo and
finally $G_0$ is the one related to $\widehat{O2}$. The last one is interpreted as a massive D2-brane
considered previously.

\item
In the case of $O2^+-$ , $G_0$ is twice an integer. This means that there is not
allowed
$\IZdos$-fluxes in K-theory, and a $O2^+$ is equivalent to an \RPOtwo-plane,
but
massive D2-branes are still present besides the usual BPS D2-branes.
\item
For $O1^-$ and $O0^-$ (\RNOone- and \RNOzero-planes) we have the usual D1 and
D0-branes (respectively) and the induced fractional $\frac{1}{2}$D1 and
$\frac{1}{2}$D0. Hence, we can write,
\beqa
\hbox{\RNOone}=O1^- +\frac{1}{2}D1, \\\nonumber
\hbox{\RNOzero}=O0^- +\frac{1}{2}D0.
\eeqa
\item
For $O1^+$ and $O0^+$ we also have equivalent orientifolds due to the fact that
there are surjective maps
\beqa
\begin{array}{cc}
d_3:H^2=\IZdos \rightarrow \widetilde{H}^5 =\IZdos, &\quad \hbox{for}\: O1^+\\
d_3:H^3=\IZdos \rightarrow \widetilde{H}^6 =\IZdos, &\quad \hbox{for}\: O0^+.\\
\end{array}
\eeqa
Then, $O1^+$ and \RPOone-planes are equivalent in K-theory. The same equivalence is found for $O0^+$ and
\RPOzero-planes.
\end{itemize}

By comparing these results given by K-theory with the cohomology classification of branes (which was
discussed in section 5), we conclude:
\begin{itemize} 
\item
\RNOp = $Op^- + \frac{1}{2}{\rm D}p$ for all $p$. It is represented by the integer flux $\IZ$ in K-theory. (Of course, this
is obtained by the use of the AHSS). 
\item $\widehat{Op}^- = Op^- + ``{\rm F}p$'' for $p <2$. ``F$p$'' is represented by the K-theory flux
$\IZdos$. 
\item $\widehat{\widetilde{Op}}^- = Op^- + \frac{1}{2}\rm{D}p + ``{\rm F}p$''. Hence, the K-theory flux $\IZ \oplus \IZdos$
represents the existence of these type of orientifold planes for $p<2$. For $p>2$ we only have a $\IZ$ charge
and this means that there is only one possibility of constructing an exotic orientifold: the \RNOp -plane. 
\end{itemize}

For the $Op^+$-plane, we have: 
\begin{itemize} 
\item 
\RPOp $=\,Op^+\,+\,``{\rm F}p$'' for $6>p>3$. With ``F$p$'' being the brane
obtained at cohomology level by the normal or twisted group, $H^{6-p}$. This is represented in 
K-theory by the flux $\IZ\oplus\IZdos$. 
\item
\RPOp $=\,Op^+$ for $p\leq3$. Although there is a cohomology group related to a $p$-brane, this is
lifted to a zero class in K-theory trough the differential map $d_3: H^{2-p}\rightarrow H^{5-p}$. Then,
the possible ``F$p$''-brane is classified in K-theory by the zero class. These orientifolds are
equivalent.  \item $\widehat{Op}^+\,=\,Op^+\,+\,``{\rm F}p"$ does not exists for $p<2$. This is because
the cohomology class $G_{2-p}$ related to the $p$-brane is obstructed to be lifted to K-theory (again,
by the presence of the non-trivial map $d_3$). Hence, $``{\rm F}p"$-brane, related to $H^{2-p}$ is not
classified by K-theory. In this sense, it is physically absent. 

\end{itemize}

Now, by applying the AHSS to the case for $d$-branes on top of $Op$-planes with $p\leq 2$, we get the results given in table \ref{p<2}.

\begin{table}
\begin{center}
\caption{Brane states on top of $Op$-planes considering discrete RR fields, $p\le 
2$.}
\label{p<2}
\begin{tabular}{||c|c|c||}\hline\hline
$O2^-$&$O1^-$&$O0^-$\\\hline\hline
D2+$^m$D2+$\frac{1}{2}$D2&-&-\\
\Done + $``{\rm F}1"$&D1+$``{\rm F}1"$ +$\frac{1}{2}$D1&-\\
\Dzero +$``{\rm F}0"$&\Dzero + $``{\rm F}0"$& D0+ $``{\rm F}0"$ + $\frac{1}{2}$D0\\
-&\Dmone&\Dmone\\\hline\hline\hline
$O2^+$&$O1^+$&$O0^+$\\\hline\hline
D2 +$^m$D2&D1&D0\\\hline\hline
\end{tabular}
\end{center}
\end{table}

\subsubsection{Equivalent and Unexistent branes}
From the point of view of cohomology, since we have two different cohomology groups associated to the
same kind of branes, we can construct two different kind of orientifold planes. The existence of this
planes depends if the relevant fluxes can be lifted to non-trivial classes in K-theory. For $Op^-$ we
have that always, the cohomology groups related to $d$-branes can be lifted to non-trivial elements
in K-theory, and that is why we have three different kinds of $Op^-$-planes.

For the $Op^+$-plane the situation is different since there exists a non-trivial map $d_3$ which obstructs
any lifting of $G_{2-p}$ fluxes to K-theory. This is the reason that
$\widehat{O1}^+$,$\widehat{O0}^+$, $\widetilde{\widehat{O1}}^+$ and $\widetilde{\widehat{O0}}^+$ do
not exist. By extending this argument to all the possible $d$-branes on top of orientifold planes, we
have that the following branes do not exist: 
\beqa 
\begin{array}{cc} 
O5^+ :&\quad \hbox{D2}\\ 
O4^+:&\quad \hbox{D2 , \Done}\\ 
O3^+ :&\quad \hbox{D2 , \Done , \Dzero}, 
\end{array} 
\eeqa 
and the following ones are represented by zero class in K-theory, 
\beqa 
\begin{array}{cc} 
O5^+:&\quad \hbox{\Dthree, \Dmone}\\ 
O4^+ :&\quad \hbox{\Dthree, \Done, \Dmone}\\ 
O3^+ :&\quad \hbox{\Dthree, \Dtwo, \Done }. 
\end{array} 
\eeqa

The same effect is observed for $d$-branes on top of $Op^+$-planes.
Consider for instance the 1-brane on top of an $O2^+-$plane. From cohomology
there
are two sources for possible $\widehat{1}$-branes. However, one of them, the $G_5$-flux is
lifted to a trivial element in K-theory, through the surjective map $d_3: H^2
=\IZdos \rightarrow \widetilde{H}^5=\IZdos\:$.
The other one, the $G_1 \in \widetilde{H}^1 =\IZdos$ flux is obstructed to be lifted
to K-theory because the map $d_3: \widetilde{H}^1 \rightarrow H^4$. Then there are
no
possible RR fluxes captured by K-theory for the $O2^+-$plane. The same
happen
with $d$-branes with $d<p$ and for $p\le 2$ for the $Op^+-$plane. This is shown in
table \ref{obs}.

The conclusion is that we can deduce the existence (or not existence) of certain  D-branes by
computing first their cohomology group and then lifting their classes to K-theory
by the $d_3$ differential map. In find that some branes do not exist even though cohomology suggests their existence.

Furthermore, with the help of the AHSS we can explain the differences between
cohomology
and K-theory. For instance, the possible \Dtwo-branes in table
\ref{coh2} do not exist for $O4^+-$plane. This is because this brane is equivalent to
vacuum
in K-theory. The \Done$\,$ brane does not exist for $O4^+$ because its respective
flux
is obstructed. We show in table \ref{obs}, all the states that are equivalent to vacuum for
$d<p$. For $d=p$ it was obtained that $Op^+$ is equivalent to \RPOp. Also we
show obstructed states on $Op^+$ and their corresponding maps.

\begin{table}
\begin{center}
\caption{Equivalent-vacuum  and obstructed branes in K-theory on top of
$Op^+-$planes}
\label{obs}
\begin{tabular}{||c|c|c|c||}\hline\hline
Orientifold&$d_3$ map&Branes $\cong$ vacuum& Unexistent branes\\\hline\hline
O5&$H^0 \rightarrow \widetilde{H}^3$&\Dthree \hspace{0.2cm} \Dmone&D2\\\hline
O4&$\widetilde{H}\rightarrow H^4$&\Dtwo&\Done\\
&$H^0\rightarrow H^3$&\Dthree \hspace{.2cm} \Dmone&D2\\\hline
03&$H^2\rightarrow \widetilde{H}^5$&\Done&\\
&$H^0 \rightarrow \widetilde{H}^3$&\Dthree&D2\\
&$\widetilde{H}^1 \rightarrow H^4$&\Dzero&\\\hline
O2&$\widetilde{H}^3 \rightarrow H^6$&\Dzero&\Dmone\\
&$\widetilde{H}^1 \rightarrow H^4$&\Dtwo&\Done\\
&$H^2 \rightarrow \widetilde{H}^5$&\Done&\Dzero\\
&$H^0 \rightarrow \widetilde{H}^3$&\Dmone&\\\hline
O1&$H^2 \rightarrow \widetilde{H}^5$&\Done&\Dzero\\
&$H^0 \rightarrow \widetilde{H}^3$&\Dmone&\\
&$\widetilde{H}^1 \rightarrow H^4$&&\Done\\
&$\widetilde{H}^3 \rightarrow H^6$&\Dzero&\Dmone\\\hline
O0&$H^4 \rightarrow \widetilde{H}^7$&\Dmone&\\
&$H^2 \rightarrow \widetilde{H}^5$&&\Dzero\\
&$H^0 \rightarrow \widetilde{H}^3$&\Dmone&\\
&$\widetilde{H}^3 \rightarrow H^6$&\Dzero&\Dmone\\\hline\hline
\end{tabular}
\end{center}
\end{table}

Some important remarks are in order: firstly, there are two types
of vacuum-equivalent branes. These are branes that have zero value in K-theory,
as the \Dthree-brane on the $O5^+$-plane. They do not exist because by K-theory
their charge must be zero (added to the fact that non-zero cohomological value is
lifted to a zero one in K-theory). The other type is a brane that its
cohomological flux-value is lifted to a zero one in K-theory but is not zero
measured by K-theory groups. This give us a topological condition (by the AHSS)
about its charge. The main examples of this type of branes are the \Dmone-brane
on $O4^+-$ and $O5^+-$planes. We interpret this fact as the condition that
discrete charge must be canceled on compact spaces. Thus, we see that by
understanding the relation between cohomology and K-theory, we can give a picture
about what it is the reason that global gauge anomalies, on suitable probe branes,
should be canceled. This is the same global gauge anomaly computed at the end of
section 3.

The absence of certain branes is explained just by obstruction in
lifting cohomology classes to K-theory. There are some branes that are obstructed
and equivalent to vacuum. For them, also a K-theory computation gives a zero
flux-value.



\section{Conclusions}

In this paper we have classified RR fields by K-theory, in string backgrounds including orientifold
planes $Op^\pm$ and $d$-branes on top of them.  We consider only $d$-branes with all their coordinates
being longitudinal to the orientifold plane. In the case $d=p$ we actually recover the orientifold
classification given in Ref. \cite{bgs}.  Some of these branes turn out to be actually D$d$-branes
(sources of the RR fields classified by K-theory), but also we find that some of such RR fields are not
in fact related to a source. So, the nature of these branes is not totally clear, although we give some
arguments which allow us to think that these branes are related to the well-know fluxbranes. Our
notations of these branes, ``F$d$", stands for our limited knowledge of their nature.

On the other hand, in order to get information about the general case $d<p$, we need (in the spirit of
\cite{bgs}) a cohomology classification of RR fields in such backgrounds and the use of the AHSS. By
wrapping D$(d+n)$-branes on compact non-trivial homological $n$-cycles of the transverse space of the
$d$-brane, $\IRP^{8-p}$ we find the cohomology groups classifying RR fields in these systems. Many new
results are found when we apply the AHSS to the above both classifications. For instance, we find that
besides the expected D-branes on top of orientifold planes, actually there are more branes related to
discrete RR fluxes. Some of them turn to be fractional D$d$-branes and the other ones ``F$d$"-branes.
In fact, in the case $d=p$, the presence of these extra branes give us the two exotic types of
orientifold planes that we already knew: \RNOp and $\widehat{Op^-}$ (for $p\leq 2$).

We also show that by analyzing all possible differential $d_3$ mappings, we were able to explain the
reason why some $d$-branes (D$d$-branes and ``F$d$"-branes) do not exist for certain values of $d$ on
top of an $Op-$plane. Indeed, for the case $d=p$ this fact reproduce one result given in \cite{bgs}:
the absence of certain exotic orientifold planes, labeled as $\widehat{Op^+}$ and
$\widehat{\widetilde{Op^+}}$ for $p<2$.

Interesting enough, we also find that in the presence of an $O5^+$- the \Dmone-brane has to appear in
an even number of them, in order we have a total zero topological charge (the topological charge of
\Dmone is $\IZdos$).  Then, since this is the condition to cancel discrete charges on compact spaces
and to avoid global gauge anomalies on suitable probe branes, we conclude that this is an effect of
going from cohomology to K-theory. This is the same condition the \Dthree- and \Dfour-branes must
satisfy on presence of an $O9^+$-plane when they are placed at a point on compact spaces.

Finally we could explain (see appendix) why the ``F$d$''-brane seems at first sight to violate T-dual
relations. This is because we have to apply T-duality on the $D(d+n)$-branes wrapped on non-trivial
homological cycles. Studying the procedure carefully we can conclude that ``F$d$''-branes also satisfy
T-duality. One would wonder if these ``F$d$''-branes have some relation to the stable non-BPS states
found in Refs. \cite{dualidad0II,bergman0}.

It will be interesting to study the M-theory lifting of the states described in this paper and observe
how the correlations and obstructions given by the differential maps and the AHSS are manifested in
M-theory.

Also, it would be interesting to study a more general cycle in which we wrap 8-branes in order to pick
up RR fields for $d$-branes on top of $Op$-planes. This requires to compute more general K-theory
groups as $KR^n(\IS^{l,m})$. These kind of cycles could give rise to a more interesting non-trivial
effects, because the 8-branes could be wrapped into a ``mixture'' of the cycles considered in (4.13).

In Refs. \cite{norma,braun} it was found the correct twisted equivariant real K-theory which classify
all the brane spectrum for certain orientifold models. In our paper it was not considered many states
included in those models and it would be very interesting to find a relation of our results with those
of Ref. \cite{braun}, by wrapping D$(d+n)$-branes on $n$-cycles, but taking all possible values of $d$
(i.e., $d>p$) and by finding homology groups for more general orientifolds.


\begin{center}
{\bf Acknowledgements}
\end{center}
\hspace{0.5cm}
We would like to thank E. Barrera, G. Castillo, N. Quiroz and A. M. Uranga for very
useful suggestions and discussions. O. L.-B. also thanks A. M. Uranga for pointing out the
problem and great support. The work of H. G.-C. is partially supported by the CONACyT (M\'exico)
grant: 33951E. The work of O. L.-B. is supported by the CONACyT Graduate Fellowship: 119267.

\appendix

\section{Topological transformation of R-R and NS-NS fluxes}
In this appendix we study in detail how some of the $d$-branes obtained by wrapping D-branes on
homology cycles are actually truly D-branes on top of $Op$-planes. We use the topological
transformation mentioned in section 5.2 and some of the properties we already know for D-branes
classified by K-theory. We analyze the following interesting cases:

\begin{itemize} 

\item 
For $\widehat{{\rm D}(p-1)}$ on an $Op^-$-plane and $\widehat{{\rm D}(p-5)}$ on an
$Op^+$-plane\footnote{Although we do not know yet which branes are related to the different kind of
orientifold planes, we infer that they are T-dual versions of D-branes on top of an $O9^-$ and
$O9^+$ planes. This will be confirmed by the use of K-theory and the AHSS.}, the
relevant cohomology group is 
\beqa 
H^{7-p}(\IRP^{8-p})=\IZdos.
\eeqa 
The RR flux near to the orientifold plane is $G_{7-p}\in\IZdos$. Now, consider a local region far
away from the orientifold plane.  In such a region, the local theory is the Type II theory. There,
$H_{NS}$ and $G_{7-p}$ are trivial forms (in the cohomology sense) and coincide with those of Type
II theories.  For Type IIA(B) theory, $p$ is even(odd) and then the RR flux $G_{7-p}$ does not
exist (in both cases). This means that the $\widehat{{\rm D}(p-1)}$ brane on top of an $Op^-$-plane
cannot be separated from the orientifold plane, because far away from the orientifold, the RR flux
becomes unstable since it is the RR flux associated to an unstable non-BPS D-brane of Type IIB
theory. This RR flux eventually become stable just after the orientifold projects out the associate
tachyon \cite{sen}. The RR field $C_{p-1}$ is zero and does not couple to any D-brane (or in other
words, this brane has zero RR charge). This is precisely the T-dual version of the behavior of the
non-BPS \Deight -brane in Type I theory. So, by this procedure, we confirm that relating the
cohomology group $H^{7-p}(\IRP^{8-p})$ to $\widehat{{\rm D}(p-1)}$ gives also the expected behavior
of a T-dual version of the \Deight -brane in Type I theory. The same happens for the T-dual version
of the \Dfour -brane in $USp(32)$ string theory using the magnetic dual NS-NS field $H_{(7)}$.

\item 

For the $\widehat{{\rm D}(p-2)}$-brane on top of an $Op^-$-plane and for the $\widehat{{\rm
D}(p-6)}$-brane on top of an $Op^+$-plane, we found that the relevant cohomology group is 
\beqa
H^{8-p}(\IRP^{8-p})=\IZdos, 
\eeqa 
where this group is twisted if the cohomology group classifying BPS D$p$-branes is untwisted and
vice versa.  Far away from the $Op^-$ -plane we can built the flux $H_{NS}G_{8-p}$ which couples to
the RR field $C_{p-1}$ in the form
\beqa 
\int_{M_{10}} H_{NS}G_{8-p}C_{p-1}. 
\eeqa 
For $p$ being an even number (i.e. Type IIA theory far away from the orientifold plane), the RR flux
$G_{8-p}$ is an even rank form and it does exists. The same is true for $p$ odd. Then, we are able
to separate the $\widehat{{\rm D}(p-2)}$-brane from the orientifold by transforming the product of
fluxes into branes. The flux $H_{NS}G_{8-p}$ is odd under the orientifold projection, because
according to the relation (\ref{un-twisted}), $G_{8-p}$ is even and $H_{NS}$ is odd. This means
that in both sides of the orientifold we have transformed topologically the product of fluxes into
a D$(p-2)$-brane and a $\overline{{\rm D}(p-2)}$-brane. They carry opposite charge by the above
argument or by the fact that the RR field $C_{p-1}$ (which couples the $(p-2)$-branes) is odd under
the orientifold projection.

This correspond to the T-dual version of the \Dseven -brane on Type I theory. The \Dseven -brane
can be constructed, as a pair of D7+\Dbseven ~modulo the orientifold action. In other words, the
D-seven-brane in Type IIB theory is unstable due to the tachyon in its spectrum, but stable when it
is on top of the $O9^-$-plane (the tachyon mode is removed out by the orientifold action).

Notice, that the flux $G_{8-p}$ has non-trivial discrete values when it is near from the
orientifold plane, but has trivial cohomology values when it is far away from the orientifold
plane. This reflects the fact that just on the orientifold plane, we have stable `non-BPS' branes,
but far away from it, we are able to decompose the brane into stable or unstable D-branes in Type
II theories. The same happens for the $\widehat{{\rm D}(p-6)}$-brane on top of an $Op^+$-plane as a
T-dual version of the non-BPS \Dthree -brane in $USp(n)$ string theory.

This procedure confirms again that the cohomology groups associated to the D-branes give all the
expected properties of the known branes (T-dual versions of Type I and $USp(32)$ string theories).
\item For the $\widehat{p}$-branes, the relevant cohomology group found was
\beqa
H^{6-p}(\IRP^{8-p})=\IZdos,
\eeqa
(twisted or untwisted). The flux $G_{6-p}$ near to the orientifold plane (positive or negative
type) has a non-trivial discrete value. However, far from it, it has a trivial cohomology value
corresponding to a RR flux in Type II theories. If $p$ is odd, we have the Type IIB theory, and for
$p$ even we have the Type IIA theory. Then, a product of fluxes can be built in the bulk, as
$H_{NS}G_{6-p}$ which is even under the orientifold projection because both $H_{NS}$ and $G_{6-p}$
are odd, or because the RR field coupling this product of fluxes $C_{p+1}$ is even under the
orientifold projection. This means that by using the flux $G_{6-p}$ we are able to construct a
product of fluxes which can be transformed topologically into D$p$-branes in the bulk. This is
consistent with the possibility of separating a D$p$-brane from the orientifold.  However, this kind of
branes acquire a non-trivial discrete RR charge when they are on top of the orientifold plane.  We
know by Ref. \cite{bgs} that in the case of an $Op^-$-plane this implies that we have a fractional
D$p$-brane, but the question remains open for the $Op^+$-plane until the use of K-theory. We study
this issue in the section 6. 
\end{itemize}


\section{T-duality Relations}

The RR fields not related to sources and listed by the right hand terms in tables \ref{Kop-} and
\ref{Kop+} are given by the K-theory group $KR^{d-10}(\{pt\})$ and then at first sight, it seems
that these fields do not obey T-duality rules, but they actually do.  Looking at table
\ref{coh2} we can relate them to some of the D-branes provided by cohomology.  In this appendix we
show how T-duality applied to ``the cohomology construction" explains the apparent T-duality violation
and in the process we also report some interesting relations at the cohomological level. Nevertheless, it is
required further analysis in order to obtain a realistic physical interpretation.

\subsection{Distinguishing D6 and fractional D6-branes on an $O6$-plane from
cohomology}

We found a `puzzle' when we consider two D6-branes on top of an $O6^-$-plane. If
these integer-charged branes are obtained by wrapping D8- and D6-branes on 2- and 0-cycles of
$\IRPtwo$ respectively, how can we distinguish which one is the integer charged D6-brane and which
one is the half-integer brane predicted by K-theory correlations as shown in
\cite{bgs}?

We can resolve the apparent puzzle by using T-duality. Take a D8 brane expanded along
coordinates 012345678 on an $O6^-$-plane on 0123456 coordinates. We can wrap
coordinates 78 on a 2-cycle of $\IRPtwo$ and obtain a D6-brane.
Also we can take a D6-brane on 0123456 coordinates and wrap it on a 0-cycle of
$\IRPtwo$ and obtain a D6-brane.

With the T-duality relations we are able to elucidate which of them is fractional. Take
T-duality on the 6 coordinate. This yields:
\begin{itemize}
\item
Two $O5^-$-planes on 012345
\item
A D7 brane on 01234578 that is wrapped on a 2-cycle of $\IRPthree$ (a $2$-cycle  on
$\IRPthree$ is transverse to $O5^-$, on 78 coordinates after $\IZdos$
projection). This gives a \Dfive-brane, i.e., this is the $G_1$-flux that by
the AHSS induces a half-integer shift on the $G_3$ flux that corresponds to
a half D5-brane \cite{bgs}.
Then, we found that this is precisely the D8-brane wrapped on a 2-cycle of $\IRPtwo$ which gives
the fractional D6-brane.
\item
A D5 brane on 012345 that is wrapped on a 0-cycle. This gives the usual D5-brane on
top of an $O5^--$plane.
\end{itemize}

The second point is confirmed also by T-dual processes depicted in \cite{sugimoto}.
If we want
to build an \RNOseven-brane by a T-dual transformation on the system $O6^-$+\RNOsix,
we have to divide the two objects by an odd number of D8-branes as domain walls. But
wrapping a D8 on a \RNOsix gives a half-integer shift on RR charge. Then when D8
branes shrinks to a point, that precisely is possible by the non-trivial 2-cycle
on
$\IRP^2$, a pair of $O6^-$ and \RNOsix-planes reduces to a pair of
$O6^--$planes.
Then T-dual configuration is always an $O7^-$-plane.

In other words, taking T-duality on coordinate 7, we get:
\begin{itemize}
\item
(By two $O6^-$-planes) An $O7^-$-plane.
\item
A D7 brane on 01234568 wrapped on a 1-cycle gives a 6-brane. However this 6-brane is
T-dual to the fractional D6 on an $O6^--$plane. Considering such brane, 
implies that
we have an \RNOsix $=O6^- +\frac{1}{2}$D6. But we know from Ref. \cite{sugimoto} that
this system
reduces to just $O6^--$planes. The absence of a D6-brane on an $O7^-$-plane
confirms that a D8-brane wrapped on a 2-cycle of $\IRPtwo$ (and classified
in cohomology by $H^0(\IRPtwo ,\IZ )=\IZ$) corresponds to the fractional D6-brane.
\item
A D7 brane on 01234567 wrapped on a 0-cycle. This is the usual D7 BPS brane on top
of an $O7^-$-plane.
\end{itemize}

\subsection{T-dual relations}

Looking at tables \ref{Kop-}, \ref{Kop+} and \ref{coh2} we find some curious behavior of the extra
branes classified by cohomology, and on the RR fields classified by K-theory which do not 
correspond to
the known RR charges. It seems they do not obey T-duality rules. However we will see that actually 
they obey them.
Consider a D$(q+n)$-brane wrapped on an $n$-cycle of $\IRP^{8-p}$, for $q\le p$ and $p+n \le 8-p$.
The brane has position coordinates\footnote{We are not considering all possible permutations of
$\sigma\{0123,\cdots q\} \in \{0123\cdots p\}$, but they give the same results.}
\begin{center}
D$(q+n): \quad 0,1,2,\cdots ,q,\,p+1\, ,\cdots\, ,p+n$\,.
\end{center}
We say that a $q$-brane is obtained by wrapping such a brane on an $n$-cycle,
(the suitable fraction of the
D-brane is spherical in covering space), i.e, in the $p+1, \cdots , p+n$
coordinates.

Now we can take T-duality on one of the coordinates defining the orientifold in two ways. Let $Op$
be the orientifold plane along coordinates,
\begin{center}
$Op\,: \quad 0,1, \cdots ,q, q+1, \cdots ,p$
\end{center}
and T-duality is taken on the $r$-coordinate, with $q< r<p$. Now we have
\beqa
\begin{array}{ccc}
O(p-1)&:&0,1, \cdots ,q,q+1,\cdots ,r-1,r+1,\cdots ,p\\
D(q+1+n)&:& 0,1,\cdots ,q,r,p+1,\cdots ,p+n.
\end{array}
\eeqa
If $r<q,p$, then
\beqa
\begin{array}{ccc}
O(p-1)&:&0,1, \cdots ,r-1,r+1,\cdots ,q,q+1, \cdots ,p\\
D(q-1+n)&:& 0,1,\cdots ,r-1,r+1,q,p+1,\cdots ,p+n
\end{array}
\eeqa
and it corresponds to a $(q-1)$-brane on top of an $O(p-1)$-plane when it is wrapped on an
$n$-cycle of $\IRP^{8-(p-1)}$. If we denote a D$(q+n)$-brane wrapped on a $n$-cycle as
D$(q+n)_n$-brane (that actually is a $q$-brane), then we saw that taken T-duality on some
longitudinal coordinate of the orientifold plane,
\beqa
\begin{array}{ccc}
Op&\rightarrow&O(p-1)\\
\rm{D}q_n&\rightarrow&\left\{ \rm{D}(q+1)_{n+1}\hfill\atop \rm{D}(q-1)_n \quad \right.\;,
\end{array}
\eeqa
depending of where T-duality is taken and with $q$ being the dimension of the D-brane in Type II 
theory. If $p<r<p+n$, 
\beqa
\begin{array}{ccc}
O(p+1)&:&0,1, \cdots ,p,r\\
D(q+n-1)&:& 0,1,\cdots ,q,p+1,\cdots ,r-1, r+1, \cdots ,p+n
\end{array}
\eeqa
When this brane is wrapped on an $(n-1)$-cycle of $\IRP^{8-(p+1)}$ it gives a
$q$-brane. Note again that the cycle corresponds to the transverse coordinates to
the orientifold plane.

The last case is when $p<n<r$. Hence,
\beqa
\begin{array}{ccc}
O(p+1)&:&0,1, \cdots ,p,r\\
D(q+n+1)&:& 0,1,\cdots ,q,p+1,\cdots ,p+n,r
\end{array}
\eeqa
It is obtained a $q$-brane on top of an $O(p+1)$-plane by wrapping this
D$(q+n+1)$-brane on an $n$-cycle of $\IRP^{8-(p+1)}$. This is summarized as follows,
\beqa
\begin{array}{ccc}
Op&\rightarrow&O(p+1)\\
\rm{D}q_n&\rightarrow&\left\{ \rm{D}(q-1)_{n-1}\hfill\atop \rm{D}(q+1)_{n} \quad \right.\;,
\end{array}
\eeqa

In order to illustrate the ideas, let us describe some examples. Take, for instance, the
\Done-brane on top of an $O6^-$-plane\footnote{Again, we are using our knowledge of K-theory
classification.}. From table \ref{coh2} we see that this brane is built by a D2-brane wrapping on a
1-cycle of $\IRPtwo$, or according to our notation, a D$2_1$-brane. The array is
\beqa
\begin{array}{ccc}
O6^-&:&\quad 0123456\\
D2&:&\quad 017
\end{array}
\eeqa
After taking T-duality on some longitudinal coordinate to the orientifold (excepting the coordinate 1), the brane corresponds 
to a D3-brane wrapping a 2-cycle of $\IRPthree$, or a D$3_2$-brane on an $O5^-$-plane. Again,
according to table \ref{coh2}, this gives a $\widehat{1}$-brane. Then by using T-duality
\begin{center}
$\widehat{1}$ (on six dimensions) $\longleftrightarrow$ $\widehat{1}$ (on five dimensions).
\end{center}
But if T-duality is taken on coordinate 1 then
\beqa
\begin{array}{ccc}
O5^-&:&023456\\
D1&:&07.
\end{array}
\eeqa
This is a D1-brane wrapped on  1-cycle of $\IRP^3$, or a D$1_1$-brane. According to our 
previous results this is
a $\widehat{0}$ -brane. Of course we need K-theory to know to which objects are these branes related
to $Op^\pm-$planes.

We conclude that,
\begin{itemize}
\item
for the well known D-branes (those classified by K-theory), the relevant T-dual connecting this
kind of branes, is:
\beqa
\begin{array}{ccccc}
Op&\rightarrow&O(p-1)\quad :
\rm{D}q_n&\rightarrow&\rm{D}(q-1)_n\;,\\
Op&\rightarrow&O(p+1)\quad :
\rm{D}q_n&\rightarrow&\rm{D}(q+1)_{n}\;.
\end{array}
\label{Tsibranas}
\eeqa
\item
The `extra' branes are related each other by the following T-dual operation:
\beqa
\begin{array}{ccccc}
Op&\rightarrow&O(p-1)\quad :
\rm{D}q_n&\rightarrow&\rm{D}(q+1)_{n+1}\;,\\
Op&\rightarrow&O(p+1)\quad :
\rm{D}q_n&\rightarrow&\rm{D}(q-1)_{n-1}\;.
\end{array}
\label{Tnobranas}
\eeqa
\end{itemize}

At the cohomological level, the RR charge seems to be not conserved, but remember that T-duality
acts over -roughly speaking- K-theory states. Looking at the tables \ref{Kop-} and
\ref{Kop+}, we see that for those fluxes not related to any source, T duality preserve the
dimension of the region where they are turned on. Thus, we conclude that for those fields,
T-duality acts as (\ref{Tnobranas}).  Certainly, it is required a more exhaustive study about
T-duality on RR fields in order to understand this behavior. We hope this remark could be useful 
in the road to elucidate this relation.

\end{document}